\PassOptionsToPackage{dvipsnames}{xcolor}
\documentclass[twocolumn]{aastex701}

\usepackage{
  graphicx,
  bm,
  amsmath,
  amssymb,
  mathtools,
  tikz,
  tikz-cd,
  mathdots,
  minted,
  tabularx,
  booktabs,
adjustbox}
\usepackage[dvipsnames]{xcolor}

\usetikzlibrary{arrows.meta,positioning,calc}

\begin{document}

\title{Shock detection in meshless finite mass hydrodynamic schemes}

\correspondingauthor{Ludwig M. B\"oss}

\author[orcid=0000-0003-4690-2774]{Ludwig M. B\"oss}
\affiliation{Department of Astronomy and Astrophysics, The University of Chicago, William Eckhart Research Center, 5640 S. Ellis Ave. Chicago, IL 60637, USA}
\email[show]{lboess@uchicago.edu}

\author[orcid=0000-0002-1687-6774]{Frederik Groth}
\affiliation{Center for Cosmology and Computational Astrophysics, Institute for Advanced Study in Physics, Zhejiang University, Hangzhou, People’s Republic of China}
\affiliation{Institute of Astronomy, School of Physics, Zhejiang University, Hangzhou 310037, People’s Republic of China}
\email[]{fgroth@usm.lmu.de}

\author[orcid=]{Klaus Dolag}
\affiliation{University Observatory, Faculty of Physics, Ludwig-Maximilians-Universität München,
            Scheinerstr. 1, Munich, 81679, Germany}
\affiliation{Max-Planck-Institute for Astrophysics,
            Karl-Schwarzschild-Straße 1, Garching, 85741, Germany}
\email[]{...}

\author[orcid=0000-0003-0939-8775]{Damiano Caprioli}
\affiliation{Department of Astronomy and Astrophysics, The University of Chicago, William Eckhart Research Center, 5640 S. Ellis Ave. Chicago, IL 60637, USA}
\email[]{}
\begin{abstract}

Shocks are prevalent in a multitude of astrophysical plasmas.
They are essential in the dissipation of gravitational potential energy during structure formation, where they channel it into heat, turbulence, magnetic fields, and particle acceleration.
Accurately capturing shocks in simulations of cosmological structure formation is a considerable challenge, due to the complexity of overlapping and colliding shocks, embedded in a medium experiencing complex shear flows. Fully Lagrangian hydrodynamical methods have historically been inferior in this regard compared to Eulerian or pseudo-Lagrangian methods. This work aims to remedy this discrepancy. We present the first efforts to implement a shockfinder algorithm with shock surface reconstruction for meshless finite mass (MFM) hydrodynamic schemes. We show its accuracy compared to analytic predictions and a legacy smoothed particle hydrodynamics (SPH)- based shock-finding method. We apply this new shock-finding method to simulations of cosmic structure formation to obtain improved predictions for shock energy dissipation and cosmic ray (CR) acceleration.

\end{abstract}

\keywords{\uat{Shocks}{2086} --- \uat{Intracluster medium}{858} --- \uat{High Energy astrophysics}{739} --- \uat{Magnetohydrodynamical simulations}{1966} --- \uat{Galaxy clusters}{584} --- \uat{Computational methods}{1965}}


\section{Introduction}

Shocks are the primary energy dissipation channel in most astrophysical systems.
In cosmic structure formation, they dissipate the potential gravitational energy which is freed in the process of hierarchical merging of substructures \citep[e.g.][for a review]{Bykov2008}. This heats the intracluster medium (ICM) to temperatures of $k_B T \approx \mathrm{keV}$ \citep[e.g.][for reviews]{boehringer+10, kravtsov+12}, injects turbulent energy into the system, which leads to the amplification of magnetic fields \citep[e.g.][for a review]{Brueggen2012}, and accelerates electrons (and likely protons) to relativistic energies \citep[e.g.][for a review]{brunetti+14}.
On smaller scales, within the interstellar medium (ISM), shocks are driven by supernova (SN) explosions, which leads to local heating of the ISM, contributing to its multi-phase nature \citep[e.g.][]{McKee1977, Joung2006, Li2015, Girichidis2016, Kim2017}, the (re-) distribution of metals \citep[e.g.][]{MacLow1999, Walch2015, Kim2020, Kolborg2022, Steinwandel2025}, and to the injection of turbulence \citep[e.g.][for reviews]{MacLow2004, Elmegreen2004}, magnetic field amplification \citep[e.g.][]{Balsara2004, Gressel2008, Rieder2016, Gent2021} and the acceleration of cosmic rays \citep[CRs, e.g.][for reviews]{blandford+87, blasi13, ruszkowski+23}.

Within (magneto-) hydrodynamic simulations of cosmic structure formation, accurate shock capturing is one of the key proving grounds of any hydro solver \citep[e.g.][]{Teyssier2002, Springel2005, Springel2010, Bryan2014, Schaller2024, opengadget3}.
On top of the accurate capture of the shock physics within the chosen hydro scheme, an additional complexity arises when trying to identify the \emph{location} of shocks within the simulation.
This is especially relevant in the highly complex simulations of cosmic structure formation, where the continuous merging of structures, active galactic nuclei (AGN) activity, or SNe drive complex, often overlapping shock waves \citep{boess+23b, lee+24a}.

Various astrophysical codes have found solutions to this problem, each tailored to the underlying hydro scheme, but sharing common design choices and yielding similar results \citep[see][for a comparison]{vazza+11}.
Most commonly, the shock capturing follows three fundamental steps:
\begin{itemize}
    \item Identify a local pressure, temperature, or density gradient, and use it to construct a potential shock normal.
    \item Calculate the local up- and downstream properties along the shock normal and evaluate the Rankine-Hugoniot (RH) conditions.
    \item Filter out spurious detections by applying a number of filters on the selected shock candidates.
\end{itemize}

This method was pioneered in Eulerian, grid-based codes.
Initial attempts identified individual shocked cells by evaluating the RH conditions between neighboring cells \citep{quilis+98, miniati+00}.
However, due to shocks being numerically broadened, this led to a bias towards higher Mach numbers in the detection, since the broadening smooths out detectable gradients.
\citet{ryu+03} improved upon this by first selecting a zone of grid cells as shock candidates based on their temperature and entropy gradients, converging velocity flows and temperature jump to neighboring cells, and then selecting the cell with the minimal local velocity divergence as the shock center. After the center was defined, they computed a Mach number along each spatial direction by comparing the temperature jump to neighboring cells and chose the maximum of these Mach numbers as the Mach number of the shock $\mathcal{M}_s = \mathrm{max}(\mathcal{M}_x, \mathcal{M}_y, \mathcal{M}_z)$ \citep[see also][]{hong+14}.

\citet{skillman+08} used a similar technique in the \textsc{Enzo} code \citep{bryan+14, oshea+04}, but introduced filtering based on an evaluation of the up- and downstream temperature and density gradients to avoid mis-detection of contact discontinuities and shear flows.
Nonetheless, the computation of a Mach number along a principal axis naturally makes their shockfinder break spherical symmetry and susceptible to projection effects. \citet{vazza+09} subsequently improved upon this for shock finding in the \textsc{Enzo} code, where they used the norm of the individual Mach number components as an improved proxy for the real Mach number.

Lagrangian hydro schemes like smoothed-particle hydrodynamics \citep[SPH; ][]{gingold+77} more naturally preserve spherical symmetry, but rely on artificial viscosity for shock capturing by adding a description for the entropy jump at a shock.
That makes the change of entropy a particle experiences a natural choice for shock detection.
This was first used for the detection of shocks in post-processing of simulation snapshots by \citet{keshet+03} and \citet{hoeft+08}.

The first implementation for on-the-fly shock detection using this method was presented by \citet{pfrommer+06}.
They derived an analytic description (supported by empirical modeling on shock tubes) for the entropy jump a SPH particle can experience per timestep as it passes through a numerically broadened shock.
This results in an equation that can be solved via root-finding methods.
To circumvent the numerical broadening, they define a decay time and record the maximum Mach number that is captured during this time interval as the sonic Mach number of the shock.

An alternative shock-finding scheme for SPH was introduced by \citet{beck+16b}, who used an SPH kernel-based approach to compute the up- and downstream quantities fed into the RH conditions.
In this method, a pressure gradient within the kernel of a particle is constructed and identified as a potential shock normal.
Neighboring particles are then projected onto this shock normal, sorted into up- and downstream according to their projected position, and weighted according to the inverted kernel of the central particle.
Subsequent projection of the particles onto two axes perpendicular to the shock normal and standard kernel weighting shapes the evaluation weight into what can be imagined as a top- and bottom-heavy cylinder.
The up- and downstream weighted quantities are then used to compute a shock speed from the RH jump conditions (if the temperature and density jumps clear high-pass filters), which together with the upstream sound speed estimate gives a value for the sonic Mach number of the shock.
However, all these SPH schemes are susceptible to numerical broadening across the SPH kernel and reliant on accurate capturing of SPH gradients \citep{hu+14, beck+16a}.

A method to leverage the benefits of both Eulerian and Lagrangian shock identification methods has been introduced for the \textsc{Arepo} code by \citet{schaal+15}.
Their moving-mesh-based hydro scheme results in a similarly narrow shock zone to the full Eulerian approach, but the pseudo-Lagrangian construction of this mesh reduces geometric effects and is not ordered according to principal geometry axes.
This makes their evaluation less prone to projection effects.
Similar to \citet{ryu+03, hong+14}, they select cells that are shock candidates based on the local velocity convergence, temperature and density gradients, as well as a high-pass filter for temperature and pressure jumps.
However, they reconstruct a shock surface by ray-tracing up the local temperature gradient to find the last cell along this path that is a shock candidate and identify this cell as the downstream.
They then ray-trace back up the temperature gradient to identify the upstream cell in the same manner.
The up- and downstream quantities are then projected onto the shock surface cell and used to evaluate the RH conditions and compute the Mach number from the temperature jump.
This results in a narrow band of shock surface cells, cleaned of numerical broadening artefacts.

This technique is not generally applicable to Lagrangian hydro schemes, due to the typically large number of neighbors in modern SPH schemes, which considerably increases the cost of the ray-tracing shock surface reconstruction.
It is, however, applicable for the meshless finite mass (MFM) hydro scheme, due to its considerably lower neighbor requirements and sharper shock structures.

Hence, this work presents our efforts to implement an \textsc{Arepo}-like shock finder in the MFM scheme of the cosmological MHD code \textsc{OpenGadget3}.
This scheme allows us to overcome historical shortcomings of shock detection in fully Lagrangian hydrodynamic schemes, closes the gap to Eulerian or pseudo-Lagrangian schemes, and provides a significant leap in the accuracy of shock detection in cosmological simulations.

We structure the paper as follows: Sec. \ref{sec:methods} contains a general description of the MFM scheme in \textsc{OpenGadget3} and a detailed description of the shock finder implementation. Within Sec. \ref{sec:numerical_tests}, we perform a suite of tests for the validity of the new scheme. In Sec. \ref{sec:cluster} we apply the new shock finder to MHD simulations of massive galaxy clusters and compare the results to the legacy SPH shock finder. Finally, Sec. \ref{sec:conclusion} contains our summary and conclusion.

\section{Methods}
\label{sec:methods}

\subsection{\textsc{OpenGadget3}}

We implemented our shock detection method in the publicly available code \textsc{OpenGadget3} \citep{opengadget3}.\footnote{The shock finder implementation will be added to the public repository once this paper has been accepted for publication.}
\textsc{OpenGadget3} is a cosmological \textsc{Tree-SPH} code based on \textsc{Gadget2} \citep{springel05}. The gravity solver uses a Barnes-Hut tree at short range and a particle-mesh (PM) grid for long-range forces.
The reference hydro scheme is a modern smoothed particle hydrodynamics (SPH) implementation \citep[][]{beck+16a} with advanced viscosity schemes for shock capturing in SPH \citep[][]{dolag+05, Cullen2010}, higher-order Wendland kernels \citep[][]{Dehnen2012}, and explicit methods for isotropic and anisotropic thermal conduction \citep[][]{Jubelgas2004, Arth2014}.

The MHD solver was presented in \citet{dolag+09b} and extended to nonideal MHD in the form of magnetic diffusion and dissipation as introduced in \citet{bonafede+11}.
Divergence of the magnetic field arising from numerical truncation errors is cleaned via the hyperbolic cleaning scheme introduced by \citet{tricco+16} (for the implementation in \textsc{OpenGadget3}, see \citet{steinwandel+25}).

 The reference shock-capturing scheme was introduced in \citet[][]{beck+16b}.

\subsection{Meshless finite mass scheme}

Our novel shock-capturing scheme uses the meshless finite mass (MFM) hydro scheme \citep{Lanson&Vila2008,Lanson&Vila2008a,Gaburov&Nitadori2011}, following the implementation introduced in \citet{groth+23}.
We outline the core concepts here, but refer the interested reader to the aforementioned paper for details.

As with the SPH scheme, MFM is fully Lagrangian.
The main difference between the methods is the hydrodynamical force calculation. While SPH computes the force based on local pressure gradients, MFM computes Riemann solver fluxes for each interacting neighbor. The effective interface area is determined based on the kernel smoothing.

Overall, this requires three steps:
First, the density and gradients of primitive variables have to be determined. These steps are shared with the SPH solver.
The gradients have then to be limited, to avoid numerical artifacts. While \textsc{OpenGadget3} includes several prescriptions for the limiting, we generally use the limiters described by \citet{Hopkins2015}.
Finally, the primitive variables are extrapolated to the effective interfaces and the Riemann solver fluxes can be computed. 

Not computing the force directly based on the gradients also avoids some of the numerical instabilities inherent to SPH, such that 32 neighbors and a cubic spline kernel \citep{Monaghan&Lattanzio1985} are sufficient for practical applications.

\subsection{On-the-fly shock finder}
\label{sec:shockfinder}
\begin{figure*}
    \centering
    \input{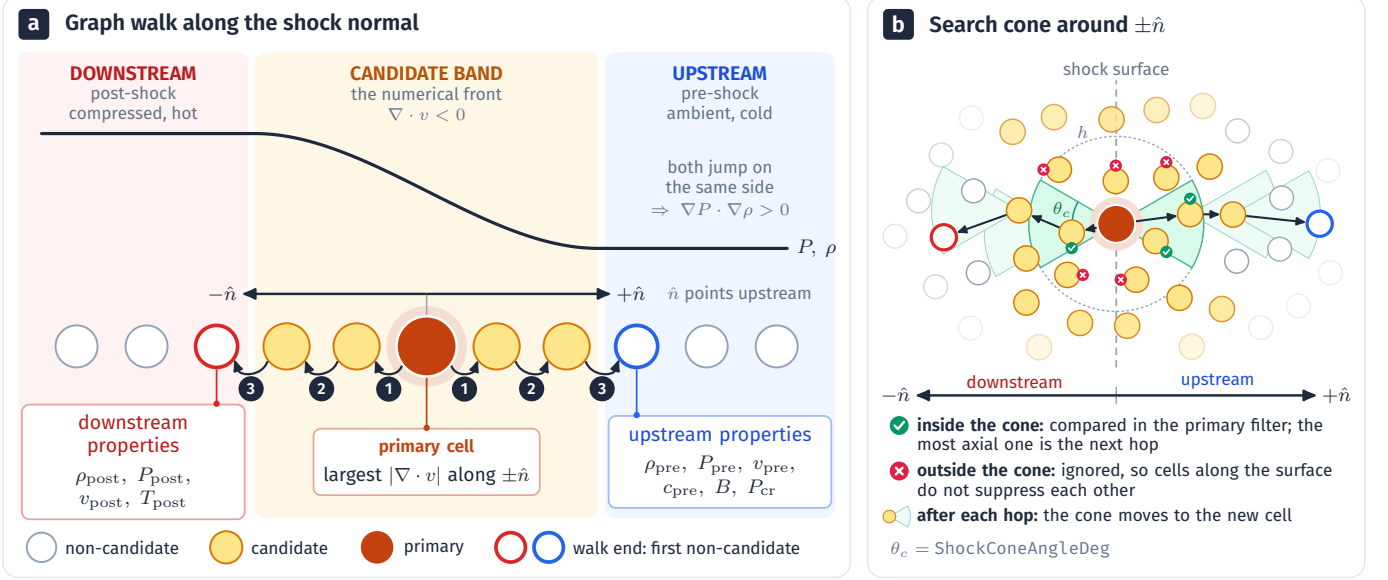}
    \caption{Schematic of the MFM shock finder. \textit{Left:} Simplified 1D view of shock surface and up-/downstream selection along the shock normal. \textit{Right:} Simplified 2D view of the up-/downstream hops along the search cones.}
    \label{fig:shockfinder_diagram}
\end{figure*}
Our shock finder implementation closely follows the methodology outlined in \citet{schaal+15}.
The algorithm is split into the five sub-steps outlined below.

\subsubsection{Shock normal construction}

The shock normal is computed following the SPH shock finder implementation described in \citet{beck+16b}.
We define the shock normal as the pressure gradient within a particle's smoothing kernel
\begin{equation}
    \mathbf{\hat{n}}_{1} = \frac{- \nabla P_i}{\vert \nabla P_i \vert}
\end{equation}
where the pressure gradient is defined as
\begin{equation}
    \nabla P_i = \frac{1}{\rho_i} \sum\limits_{j=1}^{N_\mathrm{ngb}} m_j (P_j - P_i) \: \nabla W_{ij} \: .
\end{equation}

We choose the pressure gradient, since it naturally implies a dynamic imbalance, contrary to a density or temperature gradient that could also indicate a cold front.
We use mass weighting for consistency with the SPH implementation, even though the remainder of the MFM implementation uses number weighting.
For any standard setup with equal particle masses, this does not change results.

\subsubsection{Shock candidate selection}

We flag a gas particle $i$ as a shock candidate, if it is compressive ($\nabla\!\cdot\mathbf{v} < 0$), exceeds a
floor for compression and sonic Mach number,
\begin{equation}
  \frac{|\nabla\!\cdot\mathbf{v}|\,h_i}{c_{s,i}} \;\ge\; \mathcal{M}_{\min},
  \qquad c_{s,i}=\sqrt{\gamma P_i/\rho_i},
\end{equation}
where $h_i$ is the smoothing length of particle $i$ and $c_{s,i}$ its sound speed, and passes a shear/contact-discontinuity filter
$\nabla P_i\cdot\nabla\rho_i > 0$, i.e., pressure and density increase on the same side of the potential shock surface. 
This coarse estimate only pre-selects candidates, while the Mach number is determined from the reconstructed jump, as outlined below.

\subsubsection{Primary-cell selection}
Among neighboring candidates whose normals are aligned with $i$ ($\hat{\mathbf{n}}_i\cdot\hat{\mathbf{n}}_j > \cos\theta_{\rm c}$, where $\theta_{\rm c}\!\simeq\!30^\circ$ is an empirically calibrated parameter), the cell with the largest $|\nabla\!\cdot\mathbf{v}|$ is retained as the primary cell of the shock zone.
This non-maximum suppression is applied only across the front, i.e., only to neighbors lying within a cone about $\pm\hat{\mathbf n}_i$. Tangential neighbors on the same planar surface are treated as independent fluid columns, and both remain primary candidates.

\subsubsection{Surface reconstruction}
For each primary cell, the pre-shock ($+\hat{\mathbf n}_i$) and post-shock ($-\hat{\mathbf n}_i$) states are reconstructed by following the shock normal outward until the flow leaves the shock transition.
In the moving-mesh finder of \citet{schaal+15}, a ray is propagated from cell interface to cell interface across the Voronoi tessellation.
In our meshless discretisation, we realise the same idea as a short walk over the kernel-neighbor graph.
Starting from the primary cell, each step advances a frontier $\mathbf r_{\rm f}$ to the most normal-aligned neighbor that lies ahead along $\pm\hat{\mathbf n}_i$ and within one smoothing length of the frontier,
\begin{equation}
  j^\star=\arg\max_{j}\;
  \frac{(\mathbf r_j-\mathbf r_{\rm f})\cdot(\pm\hat{\mathbf n}_i)}
       {|\mathbf r_j-\mathbf r_{\rm f}|},
  \qquad |\mathbf r_j-\mathbf r_{\rm f}| < h_i,
\end{equation}
restricted to a cone of half-angle $\theta_c=30^\circ$ about the normal. 
The walk terminates when it reaches a non-candidate particle, and this particle then defines the upstream/downstream density, pressure, velocity (and, where relevant, magnetic field and CR properties).
The shock normal is held fixed at the primary's value along the walk, and the walk is capped at $N_{\rm step}$ steps to save on computational resources.
However, each frontier particle provides its own smoothing length $h_j$ as a maximum search radius for the subsequent best-aligned shock candidate.
We find that this is necessary to capture low Mach number shocks, but can lead to a small over-prediction of the sonic Mach number for strong shocks (see Sec.~\ref{sec:numerical_tests}).

Candidates whose own normal has rotated out of the coherent front are not valid steps, which prevents the ray from crossing a contact discontinuity or leaking into an adjacent shock.

\subsubsection{Jump condition evaluation}

To reconstruct the sonic Mach number of the shock, we evaluate the Rankine-Hugoniot jump conditions across the constructed uppstream/downstream quantites of the last sub-step.

This can either be done via the temperature jump as in \citet{schaal+15},
\begin{equation}
  \frac{T_2}{T_1} =
  \frac{\bigl[2\gamma\mathcal{M}^2-(\gamma-1)\bigr]
        \bigl[(\gamma-1)\mathcal{M}^2+2\bigr]}
       {(\gamma+1)^2\,\mathcal{M}^2},
  \label{eq:tjump}
\end{equation}
with $T\propto P/\rho$ and subscripts $1,2$ denoting the up- and downstream states.
This equation can be solved in closed form, and we keep the single positive root as the Mach number estimate.
Alternatively, the Mach number can be constructed from the pressure jump via
\begin{equation}
    \frac{P_2}{P_1} = \frac{2 \gamma \mathcal{M}_s^2 - (\gamma -1)}{\gamma + 1}
\end{equation}
which can be solved for $\mathcal{M}_s$.
When including corrections due to CR feedback, as is required for the coupling to our spectral CR model, we arrive at \citep{pfrommer+17}
\begin{equation}
    \mathcal{M}_s^2 = \left( \frac{P_2}{P_1} - 1 \right) \frac{x_s}{\gamma_\mathrm{eff} (x_s - 1)} ,
    \label{eq:M_analytic}
\end{equation}
where $x_s \equiv \rho_2/\rho_1$ is the compression ratio of the shock and 
\begin{equation}
    \gamma_\mathrm{eff} \equiv \frac{\gamma_\mathrm{cr} X_\mathrm{cr} + \gamma_\mathrm{th}}{X_\mathrm{cr} + 1}
\end{equation}
is the effective adiabatic index of a mix of thermal and CR gas with $\gamma_\mathrm{cr} = 4/3$, $\gamma_\mathrm{th} = 5/3$ and $X_\mathrm{cr} = P_\mathrm{cr} / P_\mathrm{th}$.
We accept a shock only if the
reconstructed states satisfy $P_2/P_1 \ge 1.05$, $\rho_2/\rho_1 \ge 1.05$ \citep[to avoid a numerical divergence in \eqref{eq:M_analytic}, see Sec. 7.2.1 in][]{whittingham+26}, a positive
velocity jump along the normal, and $\mathcal{M}_s > 1.3$.

The shock speed is then reconstructed from the Mach number estimate and the measured upstream sound speed $v_s = c_{s,1}\,\mathcal{M}$, and the thermalisation
(dissipation) rate assigned to a cell of volume $V_i=m_i/\rho_i$ is
\begin{equation}
  \dot{E}_{\rm diss,i} =
  \tfrac{1}{2}\,\rho_1\,v_s^{3}\,\Bigl(1-\frac{1}{x_s^2}\Bigr)\,V_i^{2/3},
  \label{eq:ediss}
\end{equation}
where $V_i^{2/3}$ estimates the cell's shock-surface area.
In the absence of cooling, the reconstructed compression agrees with the analytic Rankine-Hugoniot value $r_{\rm RH}(\mathcal{M})=(\gamma+1)\mathcal{M}^2/[(\gamma-1)\mathcal{M}^2+2]$.

In magnetohydrodynamic runs we additionally record the shock obliquity, i.e. the angle between the upstream field and the shock normal,
\begin{equation}
  \theta_{Bn} = \arccos\!\left(
  \frac{|\mathbf{B}_1\cdot\hat{\mathbf n}|}{|\mathbf{B}_1|}\right)
  \in [0,\pi/2],
\end{equation}
so that quasi-parallel ($\theta_{Bn}\!\to\!0$) and quasi-perpendicular ($\theta_{Bn}\!\to\!\pi/2$) geometries are distinguished, together with the Alfv\'enic Mach number and up/downstream Alfv\'en speeds, as required by obliquity-dependent particle-acceleration models \citep{caprioli+14a, Pais2018} and non-linear DSA corrections \citep[e.g.][]{caprioli+20}.

\subsection{Timestep limiting}
To resolve fronts propagating into coarsely time-stepped gas, each shock candidate can wake any neighbor that is more than one timebin coarser to its own bin, following the limiter described by \citet{saitoh+09} and \citet{pakmor+12}.
However, in production runs we find the wake-up that is already applied to the gas particles to be sufficient to capture even sparse accretion shocks (see Sec.~\ref{sec:cluster}).
Further, in the complex shocks of cosmic structure formation, a considerable fraction of gas particles are potential shock candidates.
Lowering all of them to the same timestep floor leads to a considerable increase in computational cost for little accuracy gain \citep[see also Sec. 7.2.2 in][for a discussion of a similar problem related to the mesh construction in the \textsc{Arepo} code]{whittingham25}.

\subsection{Cosmic ray injection}

The finder localises each shock to a thin surface (of order one cell across the front), which is well suited to a clean injection of cosmic rays.
The reconstructed $v_s$, $\mathcal{M}$, $r$ and $\dot{E}_{\rm diss}$ are then used to inject into the spectral CR model, as described in \citet{boess+23a}.
The notable difference is that instead of injecting a fraction of the dissipated energy per timestep, we compute the shock crossing time through the kernel and store the time of last injection per particle to ensure that we only inject once per shock crossing, in the case where the surface reconstruction may select the same gas cell as the primary shock candidate twice in subsequent timesteps.

However, this introduces a coarser time-resolution in the CR injection and potentially leads to an under-injection in colliding shocks, which we accept in favor of a systematic over-injection when injecting at every timestep \citep[see also the discussions in][]{pfrommer+06, dubois+19}.

\section{Numerical tests\label{sec:numerical_tests}}

We present a series of numerical tests to validate our new shock-finding scheme against analytic solutions and the reference SPH shock finder by \citet{beck+16b}.

\subsection{Sod shock}
\begin{figure*}
    \centering
    \includegraphics[width=\linewidth]{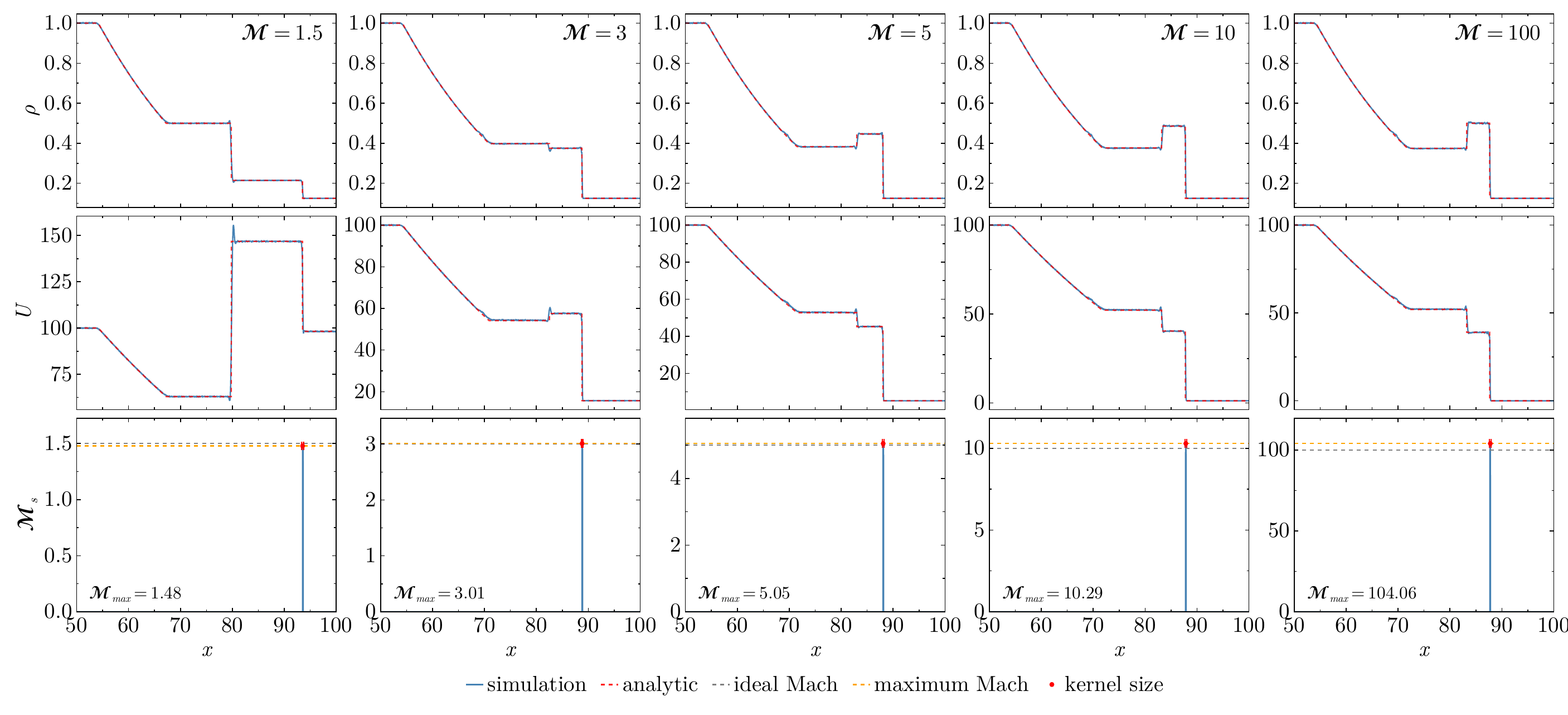}
    \caption{Suite of Sod-shock simulations in the Mach number range $\mathcal{M}_s \in [1.5, 3, 5, 10, 100]$ (from left to right). Top panels show the density, middle panels the internal energy, and bottom panels the sonic Mach number. In all panels, blue lines show the simulation output and red dashed lines the analytic solution. In the bottom panels, the analytic and maximum measured Mach numbers are indicated in the bottom left and marked by the gray dashed line for the analytic Mach number, while the orange dashed line shows the maximum measured Mach number. The red dot with error bars indicates the width of the kernel of the particle with the maximum detected Mach number.}
    \label{fig:sod_shocks}
\end{figure*}
The classic Sod shock test \citep{sod+78} is an ideal first setup for shockfinder validation, as it can be easily set to a wide range of Mach numbers.
This test is initialized as an initial density gradient $\rho_L = 8 \: \rho_R$ in a domain with size $140 \times 1 \times 1$ and periodic boundary conditions in all dimensions.
$L$ and $R$ denote the left ($x_\mathrm{min}$) and right ($x_\mathrm{max}$) side of the domain, respecting.
It is effectively a 1D test, but performed in a narrow 3D domain.
We resolve each $1 \times 1$ block in the low-density ($R$) side of the domain with $1000$ particles in a relaxed SPH glass configuration \citep{arth+19}.
The temperature on the left side of the domain is initialized to the same value in all tests, while the temperature on the right side is varied so shocks with sonic Mach number $M_s \in [1.5, 3, 5, 10, 100]$ develop.
This covers the range of typical Mach numbers for structure formation shocks found in simulations \citep{ryu+03, pfrommer+06, pfrommer+17, skillman+08, schaal+15, schaal+16}.

We show the result of this suite in Fig.~\ref{fig:sod_shocks}.
For low Mach number shocks with $\mathcal{M}_s \in [1.5, 3, 5]$, typical for merger shocks in the ICM, we find excellent agreement, with errors below the per cent level for the Mach number estimate (bottom panels).
In the high Mach number regime $\mathcal{M}_s \in [10, 100]$ associated with accretion shocks, our method overestimates the Mach number by three to five per cent.

\subsection{Weakly magnetized shock}
To test the accuracy of our shock obliquity estimate, we set up a series of Sod shocks with $\mathcal{M}_s = 10$ and a dynamically negligible background magnetic field chosen from a range of shock obliquities \citep{dubois+19, boess+23a}.
This gives us a way to quantify the quality of the upstream reconstruction, since the shock compresses the tangential component of the magnetic field, increasing the shock obliquity over the numerically broadened shock zone \citep{wittor+21}.
For this purpose, we performed a reference suite of simulations where we explicitly force $\mathbf B(\mathbf{x}, t) = \mathbf B_0(\mathbf{x})$, and with that avoid any dynamic alignment of the magnetic field with the shock surface.
This allows us to disentangle upstream reconstruction errors from the naturally expected scatter of $\theta_{B_n}$ from particle shot noise in the upstream, where the up- and downstream walk of the shock surface reconstruction can never find gas particles perfectly aligned with the shock normal.
\begin{figure}
    \centering
    \includegraphics[width=0.49\textwidth]{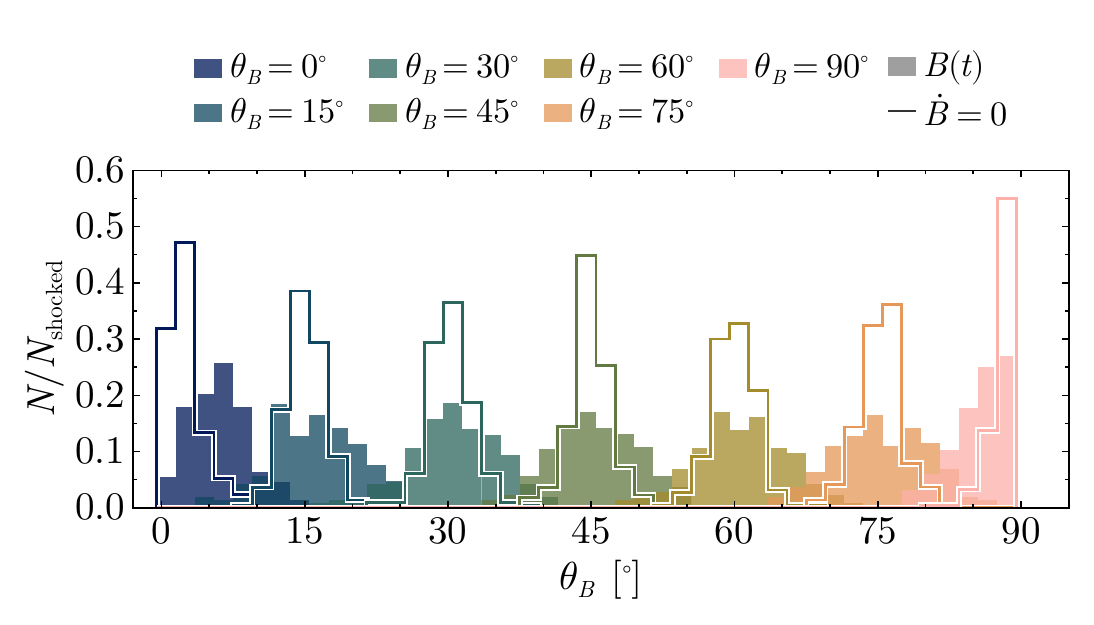}
    \caption{Normalized histograms of the measured shock obliquity. Colors correspond to the target obliquity of the run. Bars show the measured obliquity in the runs with self-consistent magnetic field evolutions; lines show the runs with a disabled magnetic field evolution.}
    \label{fig:obliquity}
\end{figure}

We show the results of these tests in Fig.~\ref{fig:obliquity}.
In general, we find reasonable agreement between the input obliquity and the measured obliquity.
The peak of the distribution aligns well with the target obliquity, with the exception of the $\theta_B = 0^\circ$ and $\theta_B = 90^\circ$ cases.
These are particularly difficult to capture due to our imposed symmetry at $\pi/2$.

When comparing to the runs with a disabled magnetic field evolution, shown as lines, we find that the width of the scatter in reconstructing $\theta_B$ reduces by roughly half.
We interpret this contribution to the reconstruction noise as particle noise in the upstream.
Since the particles do not necessarily align perfectly with the shock normal (see the right side of Fig.~\ref{fig:shockfinder_diagram}), this particle misalignment introduces noise in the angle between the shock normal and the magnetic field vector.

\subsection{Shock with CR acceleration}
\begin{figure*}
    \centering
    \includegraphics[width=0.9\linewidth]{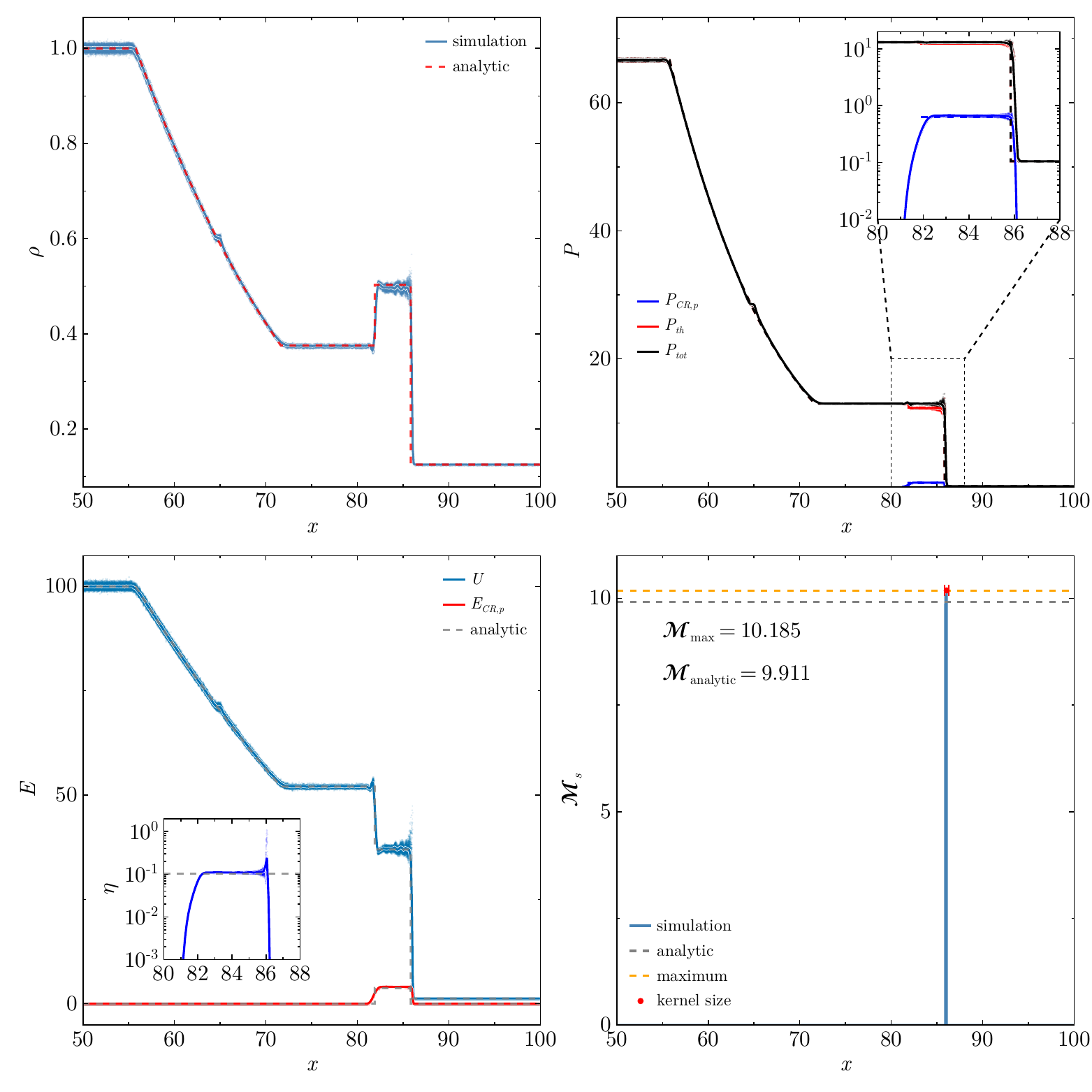}
    \caption{We show the results of a CR modified Sod shock. In each panel, the data of all particles are shown as dots and overplotted by a binned line. Dashed lines show the analytic solution. \textit{Top left:} Density. \textit{Top right:} Pressure, where blue lines show the CR proton pressure, red lines the thermal pressure, and black lines the total pressure with an inset zoom. \textit{Bottom left:} Energy, decomposed into thermal energy shown in blue and CR proton energy shown in red. The inset plot shows the acceleration efficiency. \textit{Bottom right:} Sonic Mach number with maximum measured Mach number marked with the orange dashed line, the analytic Mach number marked by the gray dashed line, and the width of the kernel of the particle with the highest Mach number marked with red error bars.}
    \label{fig:cr_sod}
\end{figure*}
We extend our tests towards a shock with CR acceleration to ensure our shock-finding scheme accurately treats the Mach number suppression of shocks with a significant CR component \citep[see][]{pfrommer+06, pfrommer+17}.

For this test, we set up a Sod shock with a target Mach number of $\mathcal{M}_s = 10$ and a constant acceleration efficiency for protons of $\eta = 0.1$.
This injected CR component modifies the shock, leading to an effective Mach number of $\mathcal{M}_s = 9.911$.

We show the results of this test in Fig.~\ref{fig:cr_sod}.
The solid lines indicate the binned result with a bin interval of $\Delta x = 0.1$ to give a smoother representation of the data, while the dots represent the output for all particles.
The tests show excellent agreement of the hydrodynamic solution and CR acceleration with the analytic solution (dashed lines).
However, the Mach number estimate has a peak that is 5 per cent above the analytic solution.
We traced this to numerical noise in the post-shock region and at the shock front, visible in the dots that represent the values of all particles.
This error is localized to part of the shock surface, while other particles reach a more accurate solution, depending on whether their up- and downstream rays end in a more or less noisy region of the shock.

We accept this small error as-is, since the uncertainty in the underlying CR acceleration efficiency is larger than this numerical error margin \citep[e.g.][]{ryu+03, caprioli+14a, guo+14a, ryu+19}.

\subsection{Sedov-Taylor blast wave}
\begin{figure}
    \centering
    \includegraphics[width=\linewidth]{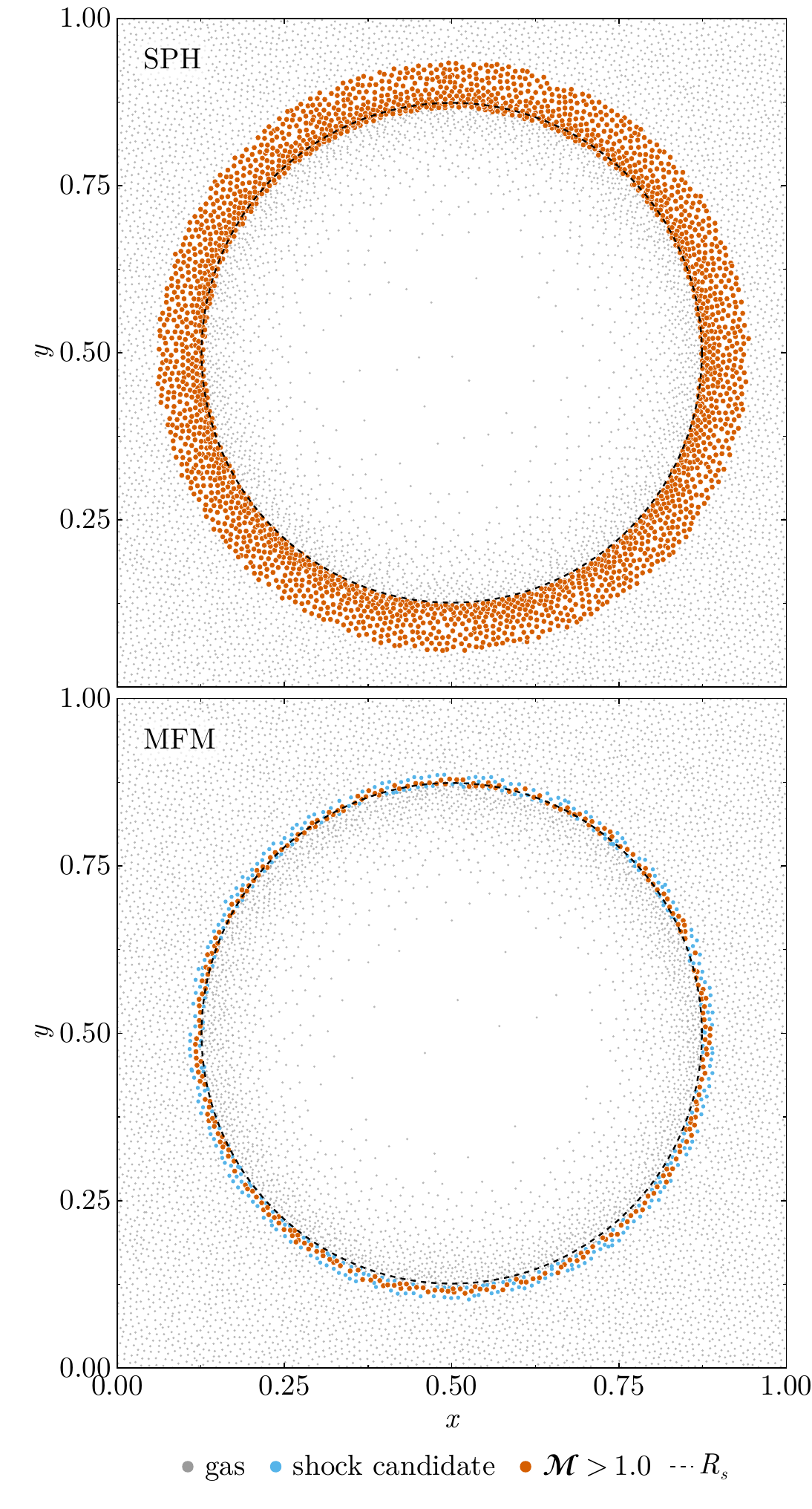}
    \caption{Comparison between the SPH and MFM shock-finding methods in a Sedov-Taylor test. We show a thin slice through the simulation domain. Gray dots indicate the position of all gas particles. Light blue dots mark the shock candidates flagged by the MFM shockfinder, and orange dots show all particles marked as shocked with a Mach number $\mathcal{M}_s > 1$. The black dashed line indicates the analytic shock radius.}
    \label{fig:sedov_candidates}
\end{figure}
\begin{figure}
    \centering
    \includegraphics[width=\linewidth]{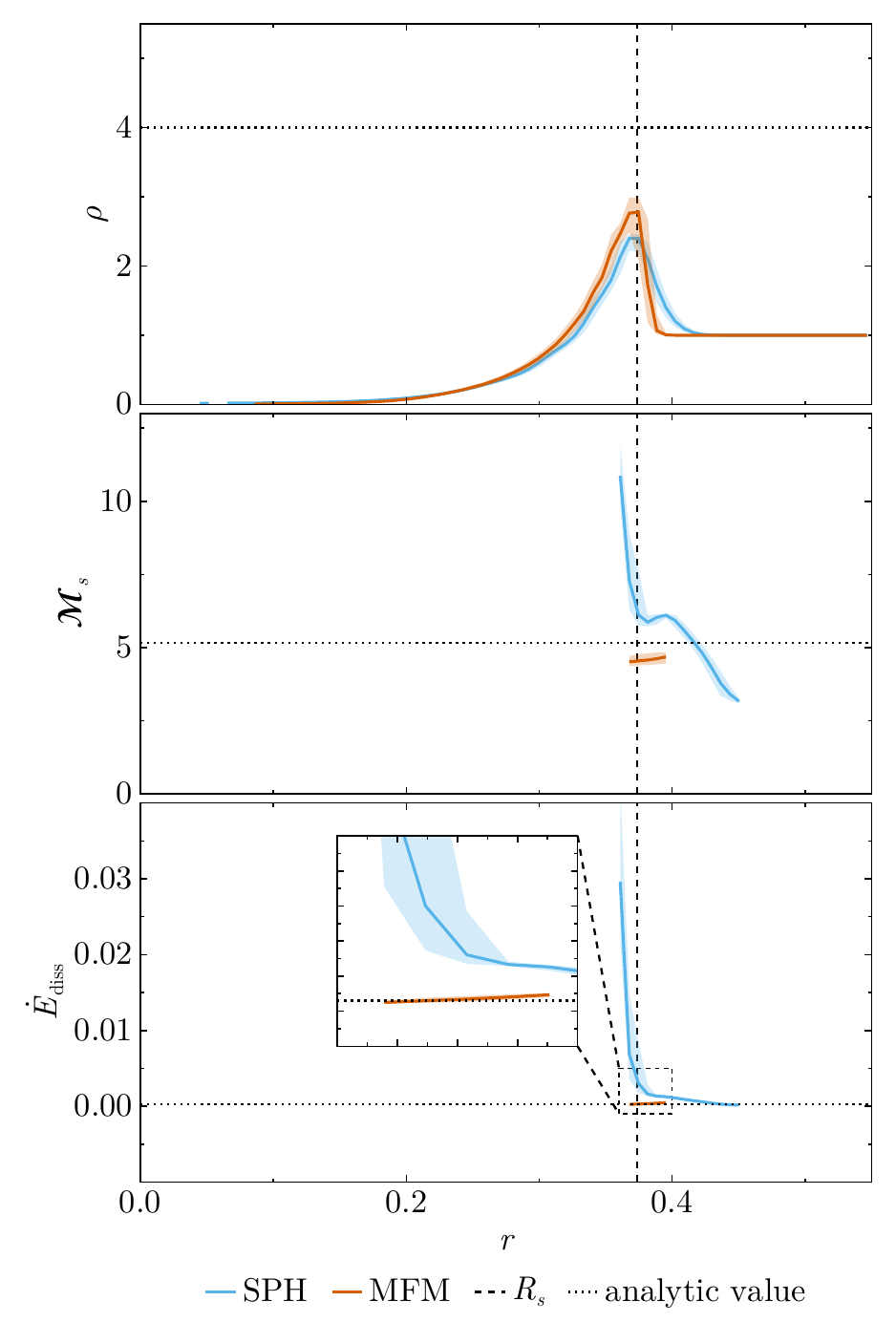}
    \caption{Radial profiles of the Sedov-Taylor test using SPH (blue) and MFM (orange). From top to bottom, we show density, sonic Mach number, and shock energy dissipation rate. Bands indicate the $1\sigma$ scatter in the radial bin. The dashed vertical line indicates the analytic position of the shock radius, and the dotted horizontal line marks the analytic solution for the respective quantity in each panel.}
    \label{fig:sedov_profiles}
\end{figure}
To depart from a simple planar shock geometry, we performed a Sedov-Taylor blastwave test.
This test mimics an idealized supernova explosion in a homogeneous background plasma, which we initialize by depositing $10^{51}$ erg in the 32 most central particles of the box.

In Fig.~\ref{fig:sedov_candidates} we illustrate the shock surface detection of our algorithm and show the reference SPH results.
The top panel shows the SPH results without surface reconstruction, where all gas cells with detected Mach numbers are marked in orange.
This indicates that, in the SPH case, the shock is smoothed out over the kernel size, and due to the symmetric inverted kernel weighting of the shock detection method and the large number of neighbors considered for the estimate, shocks are falsely detected far upstream of the shock surface.
The reason for that is upstream particles in low-density environments whose kernel overlaps with the onset of the shocked region.
When inverting the kernel for the shock detection, their large kernel size leads to them probing far into the downstream of the shock, overestimating the shock compression and, with that, the Mach number.

In contrast, the MFM shock finder benefits from a smaller number of neighbors and, with that, a smaller kernel size at a given resolution.
This results in a smaller number of shock candidates (marked as blue dots in the bottom panel) and, combined with the shock surface reconstruction method described above, a nearly single-particle surface of shocked particles, marked in orange.

This is reflected in the position of the Mach number peak relative to the analytic shock radius, marked as a dashed circle in Fig.~\ref{fig:sedov_candidates} and as a dashed horizontal line in Fig.~\ref{fig:sedov_profiles}.
While the SPH method systematically overestimates the shock radius due to the shock being detected too far upstream, the MFM method reconstructs the shock radius to excellent accuracy.

The same holds for the sonic Mach number and dissipated shock energy, as shown in the middle and bottom panel of Fig.~\ref{fig:sedov_profiles}.
While MFM shows excellent agreement, the SPH method detects a significantly broadened numerical shock.
Due to an overreach in the far downstream with the SPH method this leads to an over-estimation for Mach number, and dissipated shock energy at the inner boundary of the shock.

\section{Structure formation shocks}
\label{sec:cluster}
In order to show the improvements of our new shock-finding scheme in production simulations, we applied both our new MFM method and the legacy SPH method to a series of simulations of cosmic structure formation to show their impact on the study of shocks and CR acceleration in these environments.

\subsection{Initial conditions}

\begin{table}
    \caption{List of galaxy cluster simulations. From left to right, we list the name, mass of the gas particles, mass of the DM particles, gravitational softening of the gas particles, and the starting redshift of the respective simulation.
    \label{tab:simulations}}
    \centering
    \newlength{\tableak}
    \settowidth{\tableak}{\setlength{\tabcolsep}{0pt}\begin{tabular}{lcccc}\end{tabular}}
    \resizebox{\linewidth}{!}{%
    \hspace*{-\tableak}%
    \begin{tabular}{lcccc}
    \toprule
        Name & $M_\mathrm{gas}$ [$M_\odot$] & $M_\mathrm{DM}$ [$M_\odot$] & $\varepsilon_\mathrm{gas}$ [$h^{-1}\,c$kpc] & $z_0$\\
        \midrule
        Box4 uhr            & $1.04 \times 10^7$ & $5.2 \times 10^7$ & 0.75 & 60 \\
        Dianoga 1x          & $2.2 \times 10^8$ & $1.2 \times 10^9$ & 3.75 & 70\\
        Dianoga 10x         & $2.2 \times 10^7$ & $1.2 \times 10^8$ & 1.4   & 140 \\
        Dianoga 25x         & $8.7 \times 10^6$ & $4.7 \times 10^7$ & 1.0   & 180 \\
    \bottomrule
    \end{tabular}%
    }
\end{table}

To test our shock finder in a realistic application, we performed two classes of test simulations.
First, we performed a simulation from the Magneticum suite \citep{Dolag2025}, specifically Box4 uhr, a $50 \: h^{-1} c\mathrm{Mpc}$ box forming a galaxy group with $M_\mathrm{vir} \approx 1.6\times 10^{13} \: M_\odot$ at $z = 0$.
To save computational resources, we performed these simulations as non-radiative hydrodynamic simulations, without the inclusion of magnetic fields.
This allows us to test the shock finder in a fully cosmological context with both merger shocks and spurious accretion shocks.

We compare this to a series of zoom-in simulations at different resolution levels to test the convergence of our scheme.
As in previous works \citep{boess+23b, boess+25, steinwandel+22, steinwandel+24, diesing+26} we employ the Dianoga cluster set \citep{bonafede+11}, specifically the cluster \textsc{g5503149} at $1\times, 10\times$, and $25\times$ resolution.
We list mass and spatial resolution in Tab. \ref{tab:simulations}.
This cluster undergoes multiple violent mergers over its formation history, most notably a high impact parameter merger with a 1:10 mass ratio at $z \approx 0.3$ and a complex triple-merger at $z = 0$.
In the interest of brevity, we focus on the merger at $z = 0$.
All zoom-in simulations are nonradiative MHD simulations and were performed with SPH or MFM.
The magnetic field is initialized as constant in the $x$-direction with a strength of $B_0 = 10^{-14} \: ( 1 + z_0 )^2$ G, at the starting redshift for the simulation $z_0$.

\subsection{Shock morphology}
\begin{figure*}[t]
    \centering
    \includegraphics[width=\textwidth]{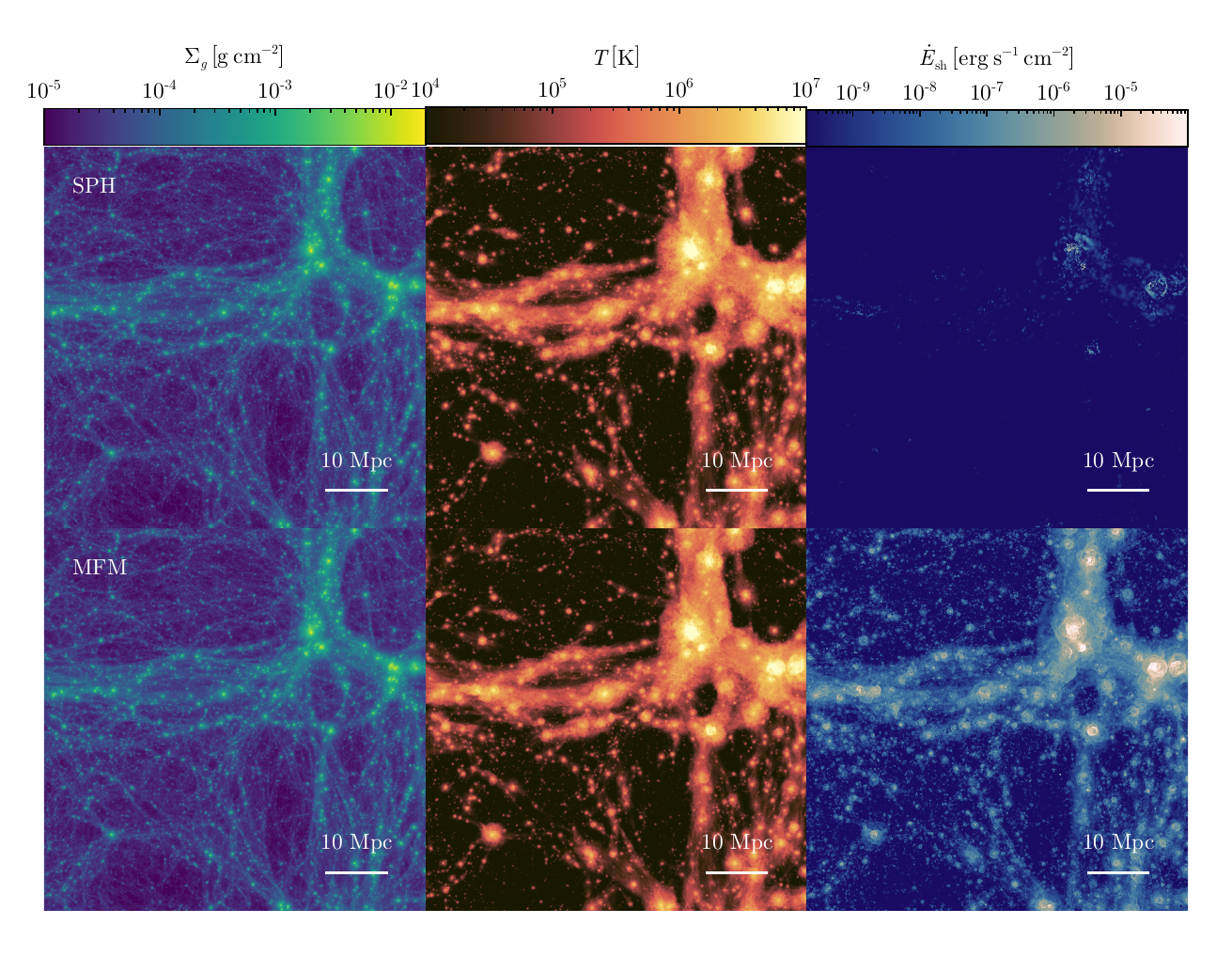}
    \caption{Projections of the \texttt{Box4 uhr} runs. From left to right, we show the surface density, temperature, and dissipated shock energy for the SPH run (top row) and the MFM run (bottom row).}
    \label{fig:box4}
\end{figure*}
\begin{figure*}[t]
    \centering
    \includegraphics[width=0.95\linewidth]{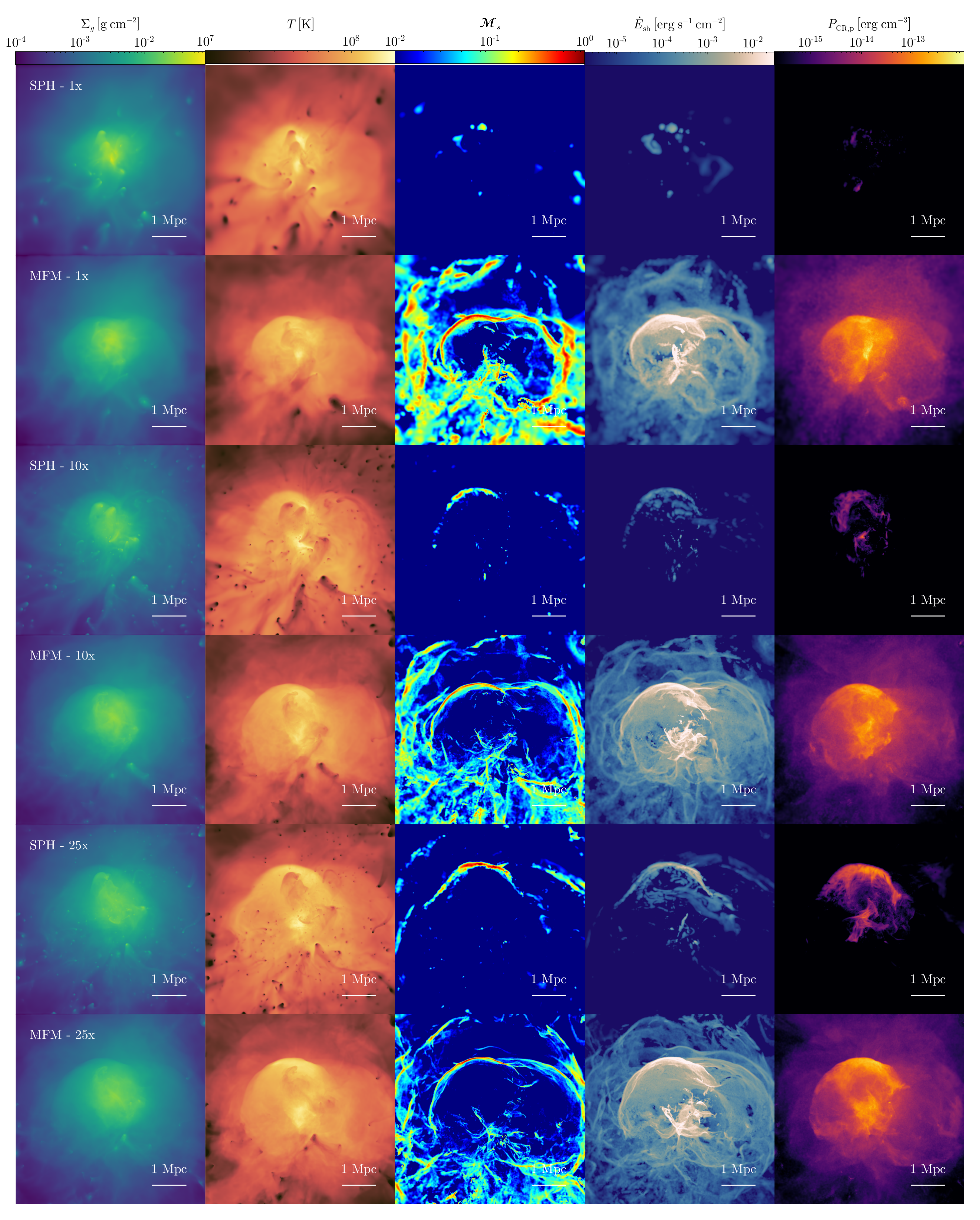}
    \caption{Comparison between resolution and hydro scheme for the galaxy cluster zoom simulations. From left to right, the columns show density, temperature, sonic Mach number, dissipated shock energy, and CR proton pressure. All quantities are shown as the mass-weighted mean along the line-of-sight. From top to bottom, we pair SPH and MFM runs at the same resolution level, indicated in the top-left of each panel, to provide a direct comparison between the methods at a fixed resolution level.}
    \label{fig:clusters}
\end{figure*}
First, we want to evaluate the performance of the shock-finding methods qualitatively.
For this, we show a comparison between the SPH and MFM runs in each setup in Figs.~\ref{fig:box4} \& \ref{fig:clusters}.

In Fig.~\ref{fig:box4} we show, from left to right, the integrated surface density, mean temperature along the line of sight and the dissipated shock energy as captured by each scheme.
For SPH, we compute the dissipated shock energy from the entropy change the particles experience (since in simulations without feedback processes shocks are the only contribution to this entropy change), while in MFM we compute the energy dissipation rate from the shock speed.
Top panels show the SPH run, while the bottom panels show the MFM results.

While the gas properties (density and temperature) agree well, the difference in capturing the dissipated shock energy is significant.
Since Box4 does not contain sufficiently large structures at $z=0$ to drive shocks with considerable energy dissipation (compared to galaxy clusters), the SPH shock finder struggles to capture most of these shocks.
We find some detection of the merger shocks in the structures of the upper-right quadrant and spurious accretion shocks detection of accretion shocks.
Comparing this to the MFM shock finder shows the significant improvement this method brings.
In the MFM runs, we find complex, layered shock structures that coincide with the merging structures in the upper-right quadrant.
We further find large-scale accretion shocks being captured by the MFM shock finder.

In Fig.~\ref{fig:clusters} we show, again from left to right, the integrated surface density, mean gas temperature, line-of-sight (LOS) averaged sonic Mach number, dissipated shock energy, and LOS-averaged CR proton pressure in the yz-plane. 
Note that we will defer the discussion of the CR proton pressure to Sec.~\ref{sec:cr_acc}.
Each set of two rows bundles a SPH run (top) and MFM run (bottom) at the same resolution level.
All plots show the final $M_\mathrm{vir} \approx 2 \times 10^{15} \: M_\odot$ cluster at $z=0$.
At this point, the cluster undergoes a merger with a small impact parameter, driving a strong shock in the z-direction.
A second, high impact parameter merger drives a weaker shock, deforming the main front of the primary shock, while a weak shock from a previous merger is still moving along the z-direction.

At 1x resolution, the SPH shock finder does not capture any of these complex structures, due to the averaging over 200 neighbors and the associated complex velocity structures within the kernel.
The MFM shock finder does not suffer from this problem due to its reconstruction of a (nearly) single-cell shock surface.
Hence, even at 1x, the MFM shock finder does recover the complex shock structures in the cluster.

Moving to 10x considerably improves the performance of the SPH shock finder; however, it still only captures the primary merger shock and only parts of the complex secondary shocks, i.e., at the previous merger shock to the right of the image in the middle panel.
In stark contrast to that, the 10x MFM run captures the complex, layered shock structures from current mergers, as well as the collision of merger- and accretion shocks \citep{zhang+20c} at the edge of the maps in the middle panel.

At 25x resolution, these trends continue.
The MFM shock finder captures the deformation of the main merger shock and more of the complex sub-shocks in the cluster periphery.
However, the advantage of the MFM shock finder is still striking, capturing the multi-layered shock structures in the cluster volume.
For the purposes of our work, and as will be shown in the following sections, we consider 10x to be converged in the MFM runs, with 25x providing further qualitative improvements.

\subsection{Shock statistics}
\begin{figure}[t]
    \centering
    \includegraphics[width=\linewidth]{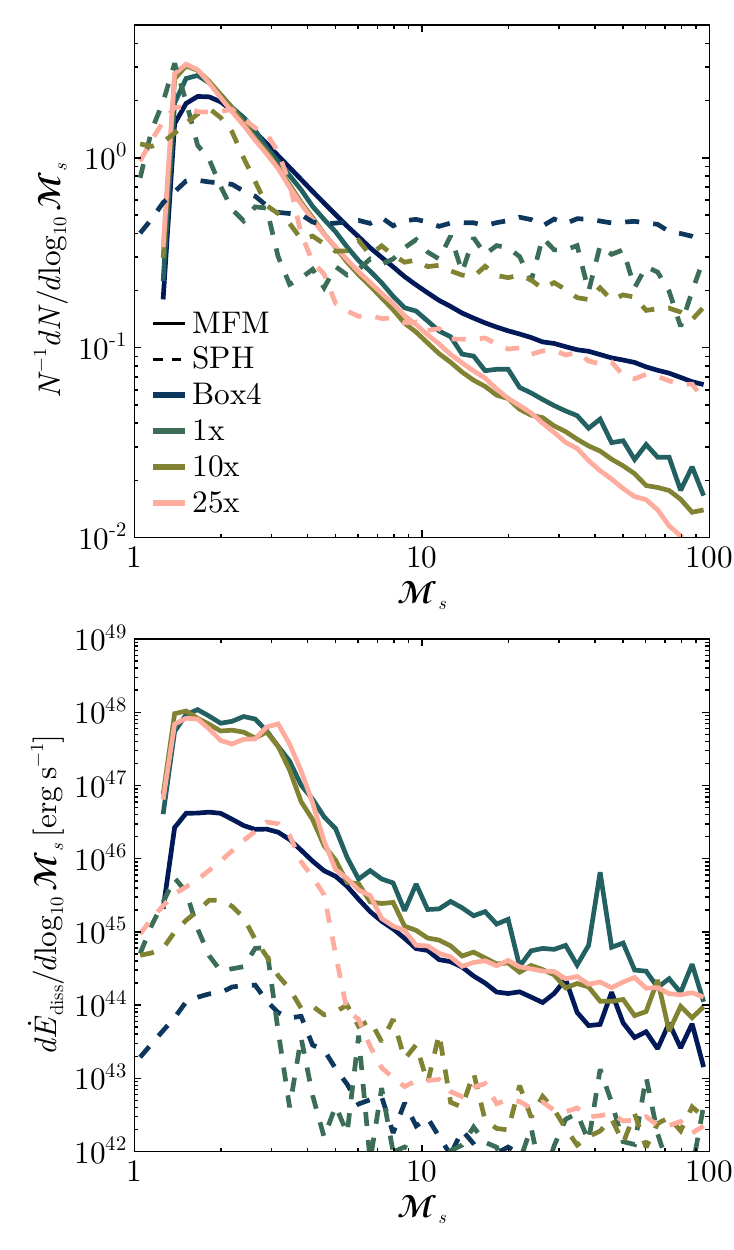}
    \caption{Histograms of shock properties. \textit{Top:} Histogram of the sonic Mach number distribution. \textit{Bottom:} Dissipated shock energy binned by sonic Mach numbers. In both panels, solid lines show the MFM runs, while dashed lines show the SPH runs. Colors correspond to the base resolution levels 1x, 10x, and 25x per run.}
    \label{fig:shock_statistics}
\end{figure}
For a more qualitative analysis, we show the histograms for the number of shocked particles and dissipated shock energy as a function of Mach number in Fig.~\ref{fig:shock_statistics}.

The top panel shows the number of shocked particles in the MFM runs (solid lines) and SPH runs (dashed lines).
Colors correspond to the Box4 simulation and the Dianoga simulations, split into resolution levels.
The histograms are normalized to the total number of shocked particles to facilitate direct comparison.
Similar to the qualitative analysis, we find the most significant difference in the Box4 uhr runs.
The SPH simulation struggles to find low Mach number shocks and predicts a significant fraction of particles to carry high Mach number shocks.
The MFM simulation provides results more in line with the statistics reported in \citet{schaal+15}: The histogram peaks at $\mathcal{M}_s \sim 1.5$ with a steep decline to $\mathcal{M}_s \sim 10$, where the decline decreases to the transition between merger- and accretion shocks.

The zoom-in simulations do not contain this considerable contribution of accretion shocks, since the high-resolution region of the simulation contains only one cluster with its attached filaments at $z=0$.
Hence, the MFM results show a similar peak at $\mathcal{M}_s \sim 1.5$, but continue their decline towards high Mach numbers, without leveling out.
Nonetheless, they show only a slight convergence trend with resolution, where higher-resolution simulations are able to capture more low Mach number shocks.
The SPH results show a significantly stronger dependence on resolution.
Similar to the qualitative analysis, it is apparent that 1x resolution is not sufficient to capture weak shocks in the $\mathcal{M}_s \sim 2-5$ regime.
While the number statistics can be considered converged between runs at 10x and 25x resolutions, in direct comparison to MFM it becomes clear that the SPH method misses shocks in the $\mathcal{M}_s \sim 3-5$ regime and overestimates the relative occurrence of high Mach number accretion shocks.

The difference between the shock-capturing schemes becomes even more significant when comparing the dissipated shock energy in the lower panel of Fig.~\ref{fig:shock_statistics}.
In the Box4 runs, the difference in captured energy dissipation across all Mach number ranges between MFM and SPH exceeds three orders of magnitude.
Hence, at this resolution, for low-mass objects, the SPH shock finder does not provide sufficient accuracy.

For the zoom-in simulations, the picture is similar, with the MFM shock finder being effectively converged already at 1x resolution level.
We attribute the remaining scatter to statistics and run-to-run variation.

In the SPH runs, we find a scatter of up-to two orders of magnitude between 1x and 25x resolution \citep[see also the discussion in][]{boess+25}.
Comparing to the MFM results allows is to conclude that at $\mathcal{M}_s \sim 3$ shocks even 25x misses one order of magnitude in dissipated energy.
For low Mach number shocks at the CR injection threshold of $\mathcal{M}_s \sim 2.3$ \citep{ha+18, ryu+19} this even approaches two orders of magnitude for 10x and 25x simulations and three orders of magnitude for 1x simulations.

\subsection{Shock obliquity}

\begin{figure}[t]
    \centering
    \includegraphics[width=\linewidth]{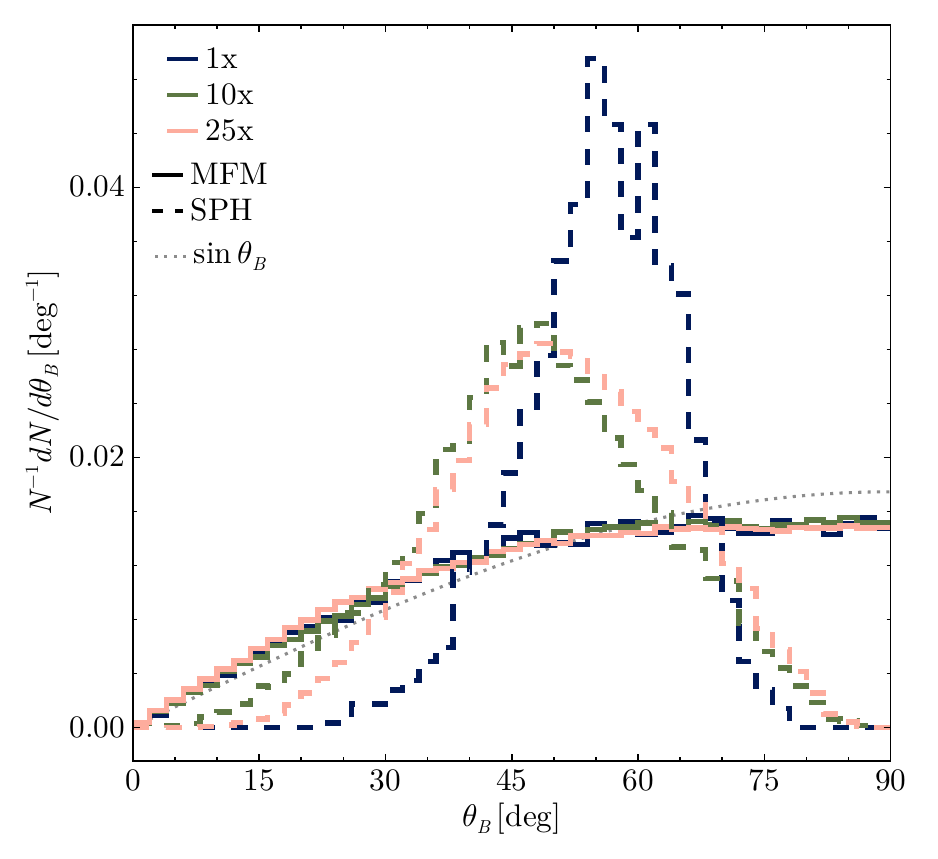}
    \caption{Histogram of shock obliquities measured in the MFM runs (solid lines) and SPH runs (dashed lines) of the Dianoga zoom-in simulations. The colors correspond to different resolution levels.}
    \label{fig:obl_hist}
\end{figure}

In the MFM runs, the obliquity distribution closely follows the isotropic expectation
$p(\theta_B)\,\mathrm{d}\theta_B = \sin\theta_B\,\mathrm{d}\theta_B$ and is converged independent of
resolution.
The SPH distributions are significantly narrower and vanish towards both $0^\circ$ and $90^\circ$.
The difference originates from how the two shock finders estimate $\theta_B$.
The MFM finder evaluates $\theta_B = \arccos|\hat{\boldsymbol{B}}\cdot \hat{\boldsymbol{n}}|$ once, with the field of the single upstream cell it selects along the shock normal $\hat{\boldsymbol{n}}$.
The SPH finder evaluates this angle for every upstream neighbor and takes the kernel-weighted mean over $\sim 200$ neighbors.
Wherever the field changes direction within the kernel, this mean of angles regresses towards the mean of the parent distribution.
A value close to $0^\circ$ or $90^\circ$ would require every neighbor to be aligned with, or perpendicular to, $\hat{\boldsymbol{n}}$.
This effect decreases with resolution in SPH, as the volume over which this average is taken decreases.

Since this effect is not present in the MFM shock finder, we can consider 1x to be converged in finding an accurate representation of the shock obliquity.

\subsection{Cosmic ray acceleration \label{sec:cr_acc}} 
\begin{figure*}[ht!]
    \centering
    \includegraphics[width=\textwidth]{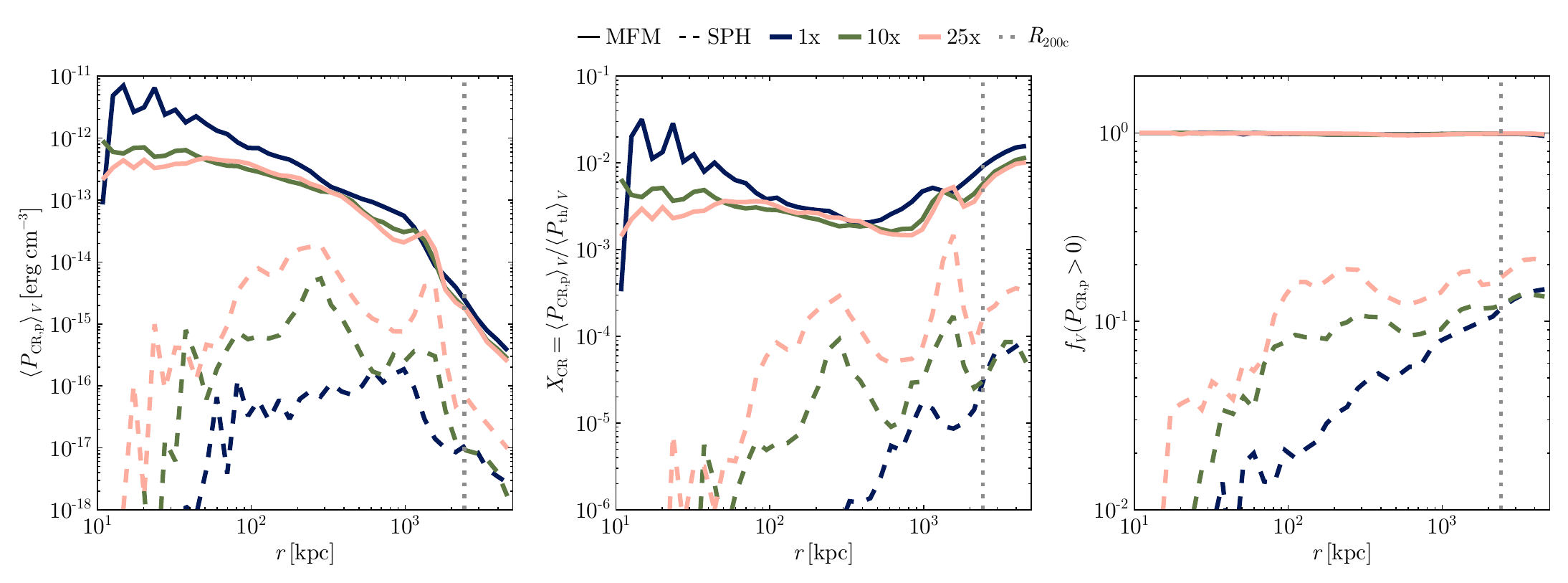}
    \caption{Radial profiles of CR proton quantities in the Dianoga zoom-in simulations. From left to right, we show the volume-averaged CR proton pressure, the ratio between thermal and CR proton pressure, and the filling fraction of CRs computed as the fraction of gas particles that contain a CR population. In all panels, solid lines show the MFM results, while dashed lines show SPH. Colors correspond to the resolution levels. The dotted line indicates $R_{200c}$ of the cluster.}
    \label{fig:crpp_radial}
\end{figure*}
We show the results of the shock-capturing scheme on our sub-grid model for CR acceleration in Fig.~\ref{fig:crpp_radial}, where we focus only on the proton component and leave the electrons for more detailed, future work.
In these panels we show from left to right: volume-averaged CR proton pressure, the volume-averaged ratio between CR proton and thermal pressure, and the volume filling fraction of gas containing a CR proton population.
As in the previous plots, solid lines mark the MFM result, while dashed lines show the SPH results and colors correspond to the resolution levels in the zoom-in simulations.
All simulations were performed using a CR acceleration efficiency of 10 per cent \citep[the canonical][result]{caprioli+14a} and used an obliquity-dependent acceleration efficiency based on recent results by \citet{gupta+25}.
Only the shock-finding scheme and resolution differ between the runs.

At the 1x resolution level, SPH and MFM differ by three to four orders of magnitude in CR pressure in the region $r > 100$ kpc, while the SPH run shows only very little CR pressure below this radius.
Compared to 10x and 25x resolution, the MFM run displays more CR pressure in the center of the cluster
Since the Mach number and energy dissipation histograms in Fig.~\ref{fig:shock_statistics} indicate convergence for MFM at all resolution levels, we attribute this to spurious over-injection of CRs at 1x resolution.

The CR proton properties indicate a clear convergence of the MFM runs at the 10x and 25x resolution levels.
All properties agree within a range we attribute to run-to-run variation.
In these runs we find a constant radial pressure ratio between CR protons and thermal gas below the per cent level required for a non-detection of $\gamma$-rays \citep{pinzke+10, ackermann+14} and a volume filling fraction of CR protons of almost 100 per cent in the entire cluster volume.
In the SPH scheme, the runs are far from converged, with increasing resolution only improving the accuracy of the CR injection at the ongoing merger shock
However, a sufficient volume-filling capture of CR acceleration cannot be achieved, leading to under-prediction of the $\gamma$-ray emission as discussed in \citet{boess+25}.

We further refer to the right panels of Fig.~\ref{fig:clusters} for a visualization of the differences in CR acceleration and the CR proton pressure in the simulated cluster volume.

\section{Conclusions} \label{sec:conclusion}

In this work, we introduced a shock-finding scheme for meshless finite mass (MFM) hydro-solvers, based on the ray-tracing scheme introduced by \citet{schaal+15} for the \texttt{Arepo} code.
In this scheme, a potential shock normal is constructed from a local pressure gradient, and shock candidates are selected from particles with a local velocity convergence above a given threshold.
We perform a walk up- and downstream of each shock candidate to check if the source particle is the particle with the highest velocity convergence along the pressure gradient.
The walk continues until either a particle that is not a shock candidate is touched, which is stored as potential up- or downstream, or a maximum number of walk steps is performed.
The particle with the maximum velocity convergence along the normal is selected to be the shock surface particle, and its up- and downstream quantities are selected to evaluate the RH jump conditions and construct a Mach number from the pressure jump.

We performed a number of numerical tests and found excellent agreement with the analytic solutions.
The tests showed good accuracy in capturing the sonic Mach number of shocks, the energy dissipation, shock surface reconstruction, CR acceleration and capturing the shock obliquity.

In simulations of cosmic structure formation, we find that the novel scheme outperforms the legacy SPH shock finder significantly, leading to numerical convergence at significantly lower resolution, and improving the potential to accurately represent CR acceleration in cosmological simulations.

\begin{acknowledgments}
L.M.B. and D.C. are partially supported by NASA through grant 80NSSC24K0173 and NSF through grant AST-2510951.
K.D. acknowledges support by the Deutsche Forschungsgemeinschaft (DFG, German Research Foundation) under Germany’s Excellence Strategy - EXC-2094 - 390783311.
This research used resources of the Argonne Leadership Computing Facility, which is a U.S. DOE of Science User Facility operated under contract DE-AC02-06CH11357.
The simulations used in this work were carried out at the Leibniz Supercomputer Center (LRZ) under the projects \textit{pn39ye} and \textit{pr83li}.
\end{acknowledgments}




\software{Julia 1.13.3 \citep{Bezanson2017}, GadgetIO.jl \citep{gadgetio}, SPHtoGrid.jl \citep{sphtogrid}, Claude Opus 5.5 \citep{Anthropic_ClaudeOpus55_2026}}

\bibliography{Total, local}{}

@ARTICLE{pinzke+10,
       author = {{Pinzke}, Anders and {Pfrommer}, Christoph},
        title = "{Simulating the {\ensuremath{\gamma}}-ray emission from galaxy clusters: a universal cosmic ray spectrum and spatial distribution}",
      journal = {\mnras},
         year = 2010,
        month = dec,
       volume = {409},
       number = {2},
        pages = {449-480},
          doi = {10.1111/j.1365-2966.2010.17328.x},
archivePrefix = {arXiv},
       eprint = {1001.5023},
 primaryClass = {astro-ph.CO},
       adsurl = {https://ui.adsabs.harvard.edu/abs/2010MNRAS.409..449P}
}

@article{caprioli+14a,
	Adsurl = {http://adsabs.harvard.edu/abs/2014ApJ...783...91C},
	Archiveprefix = {arXiv},
	Author = {{Caprioli}, D. and {Spitkovsky}, A.},
	Doi = {10.1088/0004-637X/783/2/91},
	Eid = {91},
	Eprint = {1310.2943},
	Journal = {\apj},
	Month = mar,
	Pages = {91},
	Primaryclass = {astro-ph.HE},
	Title = {{Simulations of Ion Acceleration at Non-relativistic Shocks: I. Acceleration Efficiency}},
	Volume = {783},
	Year = {2014}}

@article{blandford+87,
	Adsurl = {http://adsabs.harvard.edu/abs/1987PhR...154....1B},
	Author = {{Blandford}, R. and {Eichler}, D.},
	Doi = {10.1016/0370-1573(87)90134-7},
	Journal = {\physrep},
	Month = oct,
	Pages = {1-75},
	Title = {{Particle acceleration at astrophysical shocks: A theory of cosmic ray origin}},
	Volume = {154},
	Year = {1987}}

@article{brunetti+14,
	Archiveprefix = {arXiv},
	Author = {Brunetti, G. and Jones, T. W.},
	Doi = {10.1142/S0218271814300079},
	Eid = {1430007-98},
	Eprint = {1401.7519},
	Journal = {International Journal of Modern Physics D},
	Month = mar,
	Pages = {1430007-98},
	Title = {Cosmic Rays in Galaxy Clusters and Their Nonthermal Emission},
	Volume = {23},
	Year = {2014}}

@article{pfrommer+17,
	Archiveprefix = {arXiv},
	Author = {Pfrommer, C. and Pakmor, R. and Schaal, K. and Simpson, C. M. and Springel, V.},
	Doi = {10.1093/mnras/stw2941},
	Eprint = {1604.07399},
	Journal = {\mnras},
	Month = mar,
	Pages = {4500-4529},
	Title = {Simulating cosmic ray physics on a moving mesh},
	Volume = {465},
	Year = {2017}}

@article{wittor+21,
   title={Exploring the spectral properties of radio relics – I: integrated spectral index and Mach number},
   volume={506},
   ISSN={1365-2966},
   url={http://dx.doi.org/10.1093/mnras/stab1735},
   DOI={10.1093/mnras/stab1735},
   number={1},
   journal={Monthly Notices of the Royal Astronomical Society},
   publisher={Oxford University Press (OUP)},
   author={Wittor, D and Ettori, S and Vazza, F and Rajpurohit, K and Hoeft, M and Domínguez-Fernández, P},
   year={2021},
   month=jun, pages={396–414} }

@article{dubois+19,
	Adsurl = {https://ui.adsabs.harvard.edu/abs/2019arXiv190704300D},
	Archiveprefix = {arXiv},
	Author = {{Dubois}, Yohan and {Commer{\c{c}}on}, Beno{\^\i}t and {Marcowith}, Alexand re and {Brahimi}, Loann},
	Eid = {arXiv:1907.04300},
	Eprint = {1907.04300},
	Journal = {arXiv e-prints},
	Month = jul,
	Pages = {arXiv:1907.04300},
	Primaryclass = {astro-ph.GA},
	Title = {{Shock-accelerated cosmic rays and streaming instability in the adaptive mesh refinement code Ramses: methods and tests}},
	Year = {2019}}

@article{guo+14a,
	Archiveprefix = {arXiv},
	Author = {Guo, X. and Sironi, L. and Narayan, R.},
	Doi = {10.1088/0004-637X/794/2/153},
	Eid = {153},
	Eprint = {1406.5190},
	Journal = {\apj},
	Month = oct,
	Pages = {153},
	Primaryclass = {astro-ph.HE},
	Title = {Non-thermal Electron Acceleration in Low Mach Number Collisionless Shocks. I. Particle Energy Spectra and Acceleration Mechanism},
	Url = {http://adsabs.harvard.edu/abs/2014ApJ...794..153G},
	Volume = {794},
	Year = {2014}}

@article{caprioli+20,
	Adsurl = {https://ui.adsabs.harvard.edu/abs/2020ApJ...905....2C},
	Archiveprefix = {arXiv},
	Author = {{Caprioli}, Damiano and {Haggerty}, Colby C. and {Blasi}, Pasquale},
	Doi = {10.3847/1538-4357/abbe05},
	Eid = {2},
	Eprint = {2009.00007},
	Journal = {\apj},
	Month = dec,
	Number = {1},
	Pages = {2},
	Primaryclass = {astro-ph.HE},
	Title = {{Kinetic Simulations of Cosmic-Ray-modified Shocks. II. Particle Spectra}},
	Volume = {905},
	Year = 2020}

@article{dolag+05,
	doi = {10.1088/1475-7516/2005/01/009},
	url = {https://doi.org/10.1088/1475-7516/2005/01/009},
	year = 2005,
	month = {jan},
	publisher = {{IOP} Publishing},
	volume = {2005},
	number = {01},
	pages = {009--009},
	author = {Klaus Dolag and Dario Grasso and Volker Springel and Igor Tkachev},
	title = {Constrained simulations of the magnetic field in the local Universe and the propagation of ultrahigh energy cosmic rays},
	journal = {Journal of Cosmology and Astroparticle Physics}
}

@ARTICLE{blasi13,
       author = {{Blasi}, Pasquale},
        title = "{The origin of galactic cosmic rays}",
      journal = {A\&ARv},
         year = 2013,
        month = nov,
       volume = {21},
          eid = {70},
        pages = {70},
          doi = {10.1007/s00159-013-0070-7},
archivePrefix = {arXiv},
       eprint = {1311.7346},
 primaryClass = {astro-ph.HE},
       adsurl = {https://ui.adsabs.harvard.edu/abs/2013A&ARv..21...70B}
}

@article{ryu+03,
    author = "Ryu, Dongsu and Kang, Hyesung and Hallman, Eric and Jones, T. W.",
    title = "{Cosmological shock waves and their role in the large scale structure of the universe}",
    eprint = "astro-ph/0305164",
    archivePrefix = "arXiv",
    doi = "10.1086/376723",
    journal = "Astrophys. J.",
    volume = "593",
    pages = "599--610",
    year = "2003"
}

@ARTICLE{gupta+25,
       author = {{Gupta}, Siddhartha and {Caprioli}, Damiano and {Spitkovsky}, Anatoly},
        title = "{Speed-dependent Threshold for Electron Injection into Diffusive Shock Acceleration}",
      journal = {\apjl},
         year = 2025,
        month = nov,
       volume = {994},
       number = {2},
          eid = {L34},
        pages = {10},
          doi = {10.3847/1538-4357/ad7c4c},
archivePrefix = {arXiv},
       eprint = {2506.09134},
 primaryClass = {astro-ph.HE},
       adsurl = {https://ui.adsabs.harvard.edu/abs/2025arXiv250609134G}
}

@ARTICLE{groth+23,
       author = {{Groth}, Frederick and {Steinwandel}, Ulrich P. and {Valentini}, Milena and {Dolag}, Klaus},
        title = "{The cosmological simulation code OPENGADGET3 - implementation of meshless finite mass}",
      journal = {\mnras},
         year = 2023,
        month = nov,
       volume = {526},
       number = {1},
        pages = {616-644},
          doi = {10.1093/mnras/stad2717},
archivePrefix = {arXiv},
       eprint = {2301.03612},
 primaryClass = {astro-ph.IM},
       adsurl = {https://ui.adsabs.harvard.edu/abs/2023MNRAS.526..616G}
}

@ARTICLE{boess+23b,
       author = {{B{\"o}ss}, Ludwig M. and {Steinwandel}, Ulrich P. and {Dolag}, Klaus},
        title = "{A Formation Mechanism for ``Wrong Way'' Radio Relics}",
      journal = {\apjl},
         year = 2023,
        month = nov,
       volume = {957},
       number = {2},
          eid = {L16},
        pages = {L16},
          doi = {10.3847/2041-8213/ad03f7},
archivePrefix = {arXiv},
       eprint = {2309.00046},
 primaryClass = {astro-ph.HE},
       adsurl = {https://ui.adsabs.harvard.edu/abs/2023ApJ...957L..16B}
}

@ARTICLE{boess+23a,
       author = {{B{\"o}ss}, Ludwig M. and {Steinwandel}, Ulrich P. and {Dolag}, Klaus and {Lesch}, Harald},
        title = "{CRESCENDO: an on-the-fly Fokker-Planck solver for spectral cosmic rays in cosmological simulations}",
      journal = {\mnras},
         year = 2023,
        month = feb,
       volume = {519},
       number = {1},
        pages = {548-572},
          doi = {10.1093/mnras/stac3584},
archivePrefix = {arXiv},
       eprint = {2207.05087},
 primaryClass = {astro-ph.HE},
       adsurl = {https://ui.adsabs.harvard.edu/abs/2023MNRAS.519..548B}
}

@ARTICLE{ryu+19,
       author = {{Ryu}, Dongsu and {Kang}, Hyesung and {Ha}, Ji-Hoon},
        title = "{A Diffusive Shock Acceleration Model for Protons in Weak Quasi-parallel Intracluster Shocks}",
      journal = {\apj},
         year = 2019,
        month = sep,
       volume = {883},
       number = {1},
          eid = {60},
        pages = {60},
          doi = {10.3847/1538-4357/ab3a3a},
archivePrefix = {arXiv},
       eprint = {1905.04476},
 primaryClass = {astro-ph.HE},
       adsurl = {https://ui.adsabs.harvard.edu/abs/2019ApJ...883...60R}
}

@ARTICLE{ha+18,
       author = {{Ha}, Ji-Hoon and {Ryu}, Dongsu and {Kang}, Hyesung and {van Marle}, Allard Jan},
        title = "{Proton Acceleration in Weak Quasi-parallel Intracluster Shocks: Injection and Early Acceleration}",
      journal = {\apj},
         year = 2018,
        month = sep,
       volume = {864},
       number = {2},
          eid = {105},
        pages = {105},
          doi = {10.3847/1538-4357/aad634},
archivePrefix = {arXiv},
       eprint = {1807.09403},
 primaryClass = {astro-ph.HE},
       adsurl = {https://ui.adsabs.harvard.edu/abs/2018ApJ...864..105H}
}

@ARTICLE{ackermann+14,
       author = {{Ackermann}, M. and {Ajello}, M. and {Albert}, A. and {Allafort}, A. and {Atwood}, W.~B. and {Baldini}, L. and {Ballet}, J. and {Barbiellini}, G. and {Bastieri}, D. and {Bechtol}, K. and {Bellazzini}, R. and {Bloom}, E.~D. and {Bonamente}, E. and {Bottacini}, E. and {Brandt}, T.~J. and {Bregeon}, J. and {Brigida}, M. and {Bruel}, P. and {Buehler}, R. and {Buson}, S. and {Caliandro}, G.~A. and {Cameron}, R.~A. and {Caraveo}, P.~A. and {Cavazzuti}, E. and {Chaves}, R.~C.~G. and {Chiang}, J. and {Chiaro}, G. and {Ciprini}, S. and {Claus}, R. and {Cohen-Tanugi}, J. and {Conrad}, J. and {D'Ammando}, F. and {de Angelis}, A. and {de Palma}, F. and {Dermer}, C.~D. and {Digel}, S.~W. and {Drell}, P.~S. and {Drlica-Wagner}, A. and {Favuzzi}, C. and {Franckowiak}, A. and {Funk}, S. and {Fusco}, P. and {Gargano}, F. and {Gasparrini}, D. and {Germani}, S. and {Giglietto}, N. and {Giordano}, F. and {Giroletti}, M. and {Godfrey}, G. and {Gomez-Vargas}, G.~A. and {Grenier}, I.~A. and {Guiriec}, S. and {Gustafsson}, M. and {Hadasch}, D. and {Hayashida}, M. and {Hewitt}, J. and {Hughes}, R.~E. and {Jeltema}, T.~E. and {J{\'o}hannesson}, G. and {Johnson}, A.~S. and {Kamae}, T. and {Kataoka}, J. and {Kn{\"o}dlseder}, J. and {Kuss}, M. and {Lande}, J. and {Larsson}, S. and {Latronico}, L. and {Llena Garde}, M. and {Longo}, F. and {Loparco}, F. and {Lovellette}, M.~N. and {Lubrano}, P. and {Mayer}, M. and {Mazziotta}, M.~N. and {McEnery}, J.~E. and {Michelson}, P.~F. and {Mitthumsiri}, W. and {Mizuno}, T. and {Monzani}, M.~E. and {Morselli}, A. and {Moskalenko}, I.~V. and {Murgia}, S. and {Nemmen}, R. and {Nuss}, E. and {Ohsugi}, T. and {Orienti}, M. and {Orlando}, E. and {Ormes}, J.~F. and {Perkins}, J.~S. and {Pesce-Rollins}, M. and {Piron}, F. and {Pivato}, G. and {Rain{\`o}}, S. and {Rando}, R. and {Razzano}, M. and {Razzaque}, S. and {Reimer}, A. and {Reimer}, O. and {Ruan}, J. and {S{\'a}nchez-Conde}, M. and {Schulz}, A. and {Sgr{\`o}}, C. and {Siskind}, E.~J. and {Spandre}, G. and {Spinelli}, P. and {Storm}, E. and {Strong}, A.~W. and {Suson}, D.~J. and {Takahashi}, H. and {Thayer}, J.~G. and {Thayer}, J.~B. and {Thompson}, D.~J. and {Tibaldo}, L. and {Tinivella}, M. and {Torres}, D.~F. and {Troja}, E. and {Uchiyama}, Y. and {Usher}, T.~L. and {Vandenbroucke}, J. and {Vianello}, G. and {Vitale}, V. and {Winer}, B.~L. and {Wood}, K.~S. and {Zimmer}, S. and {Fermi-LAT Collaboration} and {Pinzke}, A. and {Pfrommer}, C.},
        title = "{Search for Cosmic-Ray-induced Gamma-Ray Emission in Galaxy Clusters}",
      journal = {\apj},
         year = 2014,
        month = may,
       volume = {787},
       number = {1},
          eid = {18},
        pages = {18},
          doi = {10.1088/0004-637X/787/1/18},
archivePrefix = {arXiv},
       eprint = {1308.5654},
 primaryClass = {astro-ph.HE},
       adsurl = {https://ui.adsabs.harvard.edu/abs/2014ApJ...787...18A}
}

@ARTICLE{ruszkowski+23,
       author = {{Ruszkowski}, Mateusz and {Pfrommer}, Christoph},
        title = "{Cosmic ray feedback in galaxies and galaxy clusters}",
      journal = {\aapr},
         year = 2023,
        month = dec,
       volume = {31},
       number = {1},
          eid = {4},
        pages = {4},
          doi = {10.1007/s00159-023-00149-2},
archivePrefix = {arXiv},
       eprint = {2306.03141},
 primaryClass = {astro-ph.HE},
       adsurl = {https://ui.adsabs.harvard.edu/abs/2023A&ARv..31....4R}
}

@ARTICLE{beck+16a,
       author = {{Beck}, A.~M. and {Murante}, G. and {Arth}, A. and {Remus}, R. -S. and {Teklu}, A.~F. and {Donnert}, J.~M.~F. and {Planelles}, S. and {Beck}, M.~C. and {F{\"o}rster}, P. and {Imgrund}, M. and {Dolag}, K. and {Borgani}, S.},
        title = "{An improved SPH scheme for cosmological simulations}",
      journal = {\mnras},
         year = 2016,
        month = jan,
       volume = {455},
       number = {2},
        pages = {2110-2130},
          doi = {10.1093/mnras/stv2443},
archivePrefix = {arXiv},
       eprint = {1502.07358},
 primaryClass = {astro-ph.CO},
       adsurl = {https://ui.adsabs.harvard.edu/abs/2016MNRAS.455.2110B}
}

@ARTICLE{beck+16b,
       author = {{Beck}, A.~M. and {Dolag}, K. and {Donnert}, J.~M.~F.},
        title = "{Geometrical on-the-fly shock detection in smoothed particle hydrodynamics}",
      journal = {\mnras},
         year = 2016,
        month = may,
       volume = {458},
       number = {2},
        pages = {2080-2087},
          doi = {10.1093/mnras/stw487},
archivePrefix = {arXiv},
       eprint = {1511.07444},
 primaryClass = {astro-ph.IM},
       adsurl = {https://ui.adsabs.harvard.edu/abs/2016MNRAS.458.2080B}
}

@ARTICLE{dolag+09b,
       author = {{Dolag}, K. and {Stasyszyn}, F.},
        title = "{An MHD GADGET for cosmological simulations}",
      journal = {\mnras},
         year = 2009,
        month = oct,
       volume = {398},
       number = {4},
        pages = {1678-1697},
          doi = {10.1111/j.1365-2966.2009.15181.x},
archivePrefix = {arXiv},
       eprint = {0807.3553},
 primaryClass = {astro-ph},
       adsurl = {https://ui.adsabs.harvard.edu/abs/2009MNRAS.398.1678D}
}

@ARTICLE{bonafede+11,
       author = {{Bonafede}, A. and {Dolag}, K. and {Stasyszyn}, F. and {Murante}, G. and {Borgani}, S.},
        title = "{A non-ideal magnetohydrodynamic GADGET: simulating massive galaxy clusters}",
      journal = {\mnras},
         year = 2011,
        month = dec,
       volume = {418},
       number = {4},
        pages = {2234-2250},
          doi = {10.1111/j.1365-2966.2011.19523.x},
archivePrefix = {arXiv},
       eprint = {1107.0968},
 primaryClass = {astro-ph.CO},
       adsurl = {https://ui.adsabs.harvard.edu/abs/2011MNRAS.418.2234B}
}

@ARTICLE{springel05,
       author = {{Springel}, Volker},
        title = "{The cosmological simulation code GADGET-2}",
      journal = {\mnras},
         year = 2005,
        month = dec,
       volume = {364},
       number = {4},
        pages = {1105-1134},
          doi = {10.1111/j.1365-2966.2005.09655.x},
archivePrefix = {arXiv},
       eprint = {astro-ph/0505010},
 primaryClass = {astro-ph},
       adsurl = {https://ui.adsabs.harvard.edu/abs/2005MNRAS.364.1105S}
}

@ARTICLE{tricco+16,
       author = {{Tricco}, Terrence S. and {Price}, Daniel J. and {Bate}, Matthew R.},
        title = "{Constrained hyperbolic divergence cleaning in smoothed particle magnetohydrodynamics with variable cleaning speeds}",
      journal = {Journal of Computational Physics},
         year = 2016,
        month = oct,
       volume = {322},
        pages = {326-344},
          doi = {10.1016/j.jcp.2016.06.053},
archivePrefix = {arXiv},
       eprint = {1607.02394},
 primaryClass = {astro-ph.IM},
       adsurl = {https://ui.adsabs.harvard.edu/abs/2016JCoPh.322..326T}
}

@ARTICLE{zhang+20c,
       author = {{Zhang}, Congyao and {Churazov}, Eugene and {Dolag}, Klaus and {Forman}, William R. and {Zhuravleva}, Irina},
        title = "{Encounters of merger and accretion shocks in galaxy clusters and their effects on intracluster medium}",
      journal = {\mnras},
         year = 2020,
        month = may,
       volume = {494},
       number = {3},
        pages = {4539-4547},
          doi = {10.1093/mnras/staa1013},
archivePrefix = {arXiv},
       eprint = {2001.10959},
 primaryClass = {astro-ph.GA},
       adsurl = {https://ui.adsabs.harvard.edu/abs/2020MNRAS.494.4539Z}
}

@ARTICLE{steinwandel+22,
       author = {{Steinwandel}, Ulrich P. and {B{\"o}ss}, Ludwig M. and {Dolag}, Klaus and {Lesch}, Harald},
        title = "{On the Small-scale Turbulent Dynamo in the Intracluster Medium: A Comparison to Dynamo Theory}",
      journal = {\apj},
         year = 2022,
        month = jul,
       volume = {933},
       number = {2},
          eid = {131},
        pages = {131},
          doi = {10.3847/1538-4357/ac715c},
archivePrefix = {arXiv},
       eprint = {2108.07822},
 primaryClass = {astro-ph.CO},
       adsurl = {https://ui.adsabs.harvard.edu/abs/2022ApJ...933..131S}
}

@ARTICLE{steinwandel+24,
       author = {{Steinwandel}, Ulrich P. and {Dolag}, Klaus and {B{\"o}ss}, Ludwig M. and {Marin-Gilabert}, Tirso},
        title = "{Toward Cosmological Simulations of the Magnetized Intracluster Medium with Resolved Coulomb Collision Scale}",
      journal = {\apj},
         year = 2024,
        month = jun,
       volume = {967},
       number = {2},
          eid = {125},
        pages = {125},
          doi = {10.3847/1538-4357/ad39ee},
archivePrefix = {arXiv},
       eprint = {2306.04692},
 primaryClass = {astro-ph.CO},
       adsurl = {https://ui.adsabs.harvard.edu/abs/2024ApJ...967..125S}
}

@ARTICLE{boess+25,
       author = {{B{\"o}ss}, Ludwig M. and {Khabibullin}, Ildar and {Karner}, Daniel and {Dolag}, Klaus and {Steinwandel}, Ulrich P. and {Hernandez-Martinez}, Elena and {Sorce}, Jenny G.},
        title = "{Simulating the LOcal Web (SLOW) - VI: $\gamma$-ray Emission in the Local Universe}",
      journal = {arXiv e-prints},
         year = 2025,
        month = oct,
          eid = {arXiv:2510.15634},
        pages = {arXiv:2510.15634},
archivePrefix = {arXiv},
       eprint = {2510.15634},
 primaryClass = {astro-ph.HE},
       adsurl = {https://ui.adsabs.harvard.edu/abs/2025arXiv251015634B}
}

@ARTICLE{steinwandel+25,
       author = {{Steinwandel}, Ulrich P. and {Price}, Daniel J.},
        title = "{$\nabla\cdot{B}=0$ versus the Universe}",
      journal = {arXiv e-prints},
         year = 2025,
        month = nov,
          eid = {arXiv:2511.19615},
        pages = {arXiv:2511.19615},
          doi = {10.48550/arXiv.2511.19615},
archivePrefix = {arXiv},
       eprint = {2511.19615},
 primaryClass = {astro-ph.CO},
       adsurl = {https://ui.adsabs.harvard.edu/abs/2025arXiv251119615S}
}

@ARTICLE{diesing+26,
       author = {{Diesing}, Rebecca and {B{\"o}ss}, Ludwig and {Caprioli}, Damiano},
        title = "{Shock Acceleration in the Intracluster Medium: Implications of Micromirror Confinement}",
      journal = {arXiv e-prints},
         year = 2025,
        month = oct,
          eid = {arXiv:2510.23700},
        pages = {arXiv:2510.23700},
          doi = {10.48550/arXiv.2510.23700},
archivePrefix = {arXiv},
       eprint = {2510.23700},
 primaryClass = {astro-ph.HE},
       adsurl = {https://ui.adsabs.harvard.edu/abs/2025arXiv251023700D}
}

@ARTICLE{whittingham+26,
       author = {{Whittingham}, Joseph and {Pfrommer}, Christoph and {Werhahn}, Maria and {Jlassi}, L{\'e}na and {Girichidis}, Philipp},
        title = "{Zooming in on cluster radio relics: I. How density fluctuations explain the Mach number discrepancy, microgauss magnetic fields, and spectral index variations}",
      journal = {\aap},
         year = 2026,
        month = jan,
       volume = {706},
          eid = {A39},
        pages = {A39},
          doi = {10.1051/0004-6361/202453002},
archivePrefix = {arXiv},
       eprint = {2411.11947},
 primaryClass = {astro-ph.HE},
       adsurl = {https://ui.adsabs.harvard.edu/abs/2026A&A...706A..39W}
}

@ARTICLE{opengadget3,
       author = {{Dolag}, Klaus and {Karademir}, Geray S. and {Tornatore}, Luca and {Ragagnin}, Antonio and {Valentini}, Milena and {B{\"o}ss}, Ludwig M. and {Groth}, Frederick and {Murante}, Giuseppe and {Borgani}, Stefano and {Springel}, Volker and {Arido}, Cenanda and {Bad{\'\i}a}, Javier and {Castro}, Tiago and {Damiano}, Alice and {Di Federico}, Laura and {Fischer}, Moritz S. and {Granato}, Gian Luigi and {Hariharan}, Nitya and {Klemmer}, David and {Maio}, Umberto and {Marin-Gilabert}, Tirso and {Marini}, Ilaria and {Patil}, Yashraj and {Egelhofer}, Margarita and {Ragone-Figueroa}, Cinthia and {Sala}, Luca and {Scannapieco}, Cecilia and {Schmidt}, Ludwig D. and {Steinwandel}, Ulrich P. and {Tedeschi-Prades}, Giovanni and {Wiertel}, Marc},
        title = "{The OpenGadget3 Code for Cosmological Simulations}",
      journal = {arXiv e-prints},
         year = 2026,
        month = sep,
          eid = {arXiv:2609.17658},
        pages = {arXiv:2609.17658},
          doi = {10.48550/arXiv.2609.17658},
archivePrefix = {arXiv},
       eprint = {2609.17658},
 primaryClass = {astro-ph.IM},
       adsurl = {https://ui.adsabs.harvard.edu/abs/2026arXiv260917658D}
}

@PHDTHESIS{whittingham25,
       author = {{Whittingham}, Joseph},
        title = "{Merging galaxies and clusters: Insights into the role of magnetic fields and the physics of radio relics}",
       school = {Leibniz Institute for Astrophysics, Potsdam},
         year = 2025,
        month = feb,
       adsurl = {https://ui.adsabs.harvard.edu/abs/2025PhDT.......132W}
}

@ARTICLE{gingold+77,
       author = {{Gingold}, R.~A. and {Monaghan}, J.~J.},
        title = "{Smoothed particle hydrodynamics: theory and application to non-spherical stars.}",
      journal = {\mnras},
         year = 1977,
        month = nov,
       volume = {181},
        pages = {375-389},
          doi = {10.1093/mnras/181.3.375},
       adsurl = {https://ui.adsabs.harvard.edu/abs/1977MNRAS.181..375G}
}

@ARTICLE{saitoh+09,
       author = {{Saitoh}, Takayuki R. and {Makino}, Junichiro},
        title = "{A Necessary Condition for Individual Time Steps in SPH Simulations}",
      journal = {\apjl},
         year = 2009,
        month = jun,
       volume = {697},
       number = {2},
        pages = {L99-L102},
          doi = {10.1088/0004-637X/697/2/L99},
archivePrefix = {arXiv},
       eprint = {0808.0773},
 primaryClass = {astro-ph},
       adsurl = {https://ui.adsabs.harvard.edu/abs/2009ApJ...697L..99S}
}

@ARTICLE{pakmor+12,
       author = {{Pakmor}, R. and {Edelmann}, P. and {R{\"o}pke}, F.~K. and {Hillebrandt}, W.},
        title = "{Stellar GADGET: a smoothed particle hydrodynamics code for stellar astrophysics and its application to Type Ia supernovae from white dwarf mergers}",
      journal = {\mnras},
         year = 2012,
        month = aug,
       volume = {424},
       number = {3},
        pages = {2222-2231},
          doi = {10.1111/j.1365-2966.2012.21383.x},
archivePrefix = {arXiv},
       eprint = {1205.5806},
 primaryClass = {astro-ph.HE},
       adsurl = {https://ui.adsabs.harvard.edu/abs/2012MNRAS.424.2222P}
}

@ARTICLE{kravtsov+12,
       author = {{Kravtsov}, Andrey V. and {Borgani}, Stefano},
        title = "{Formation of Galaxy Clusters}",
      journal = {\araa},
         year = 2012,
        month = sep,
       volume = {50},
        pages = {353-409},
          doi = {10.1146/annurev-astro-081811-125502},
archivePrefix = {arXiv},
       eprint = {1205.5556},
 primaryClass = {astro-ph.CO},
       adsurl = {https://ui.adsabs.harvard.edu/abs/2012ARA&A..50..353K}
}

@ARTICLE{boehringer+10,
       author = {{B{\"o}hringer}, Hans and {Werner}, Norbert},
        title = "{X-ray spectroscopy of galaxy clusters: studying astrophysical processes in the largest celestial laboratories}",
      journal = {\aapr},
         year = 2010,
        month = feb,
       volume = {18},
       number = {1-2},
        pages = {127-196},
          doi = {10.1007/s00159-009-0023-3},
       adsurl = {https://ui.adsabs.harvard.edu/abs/2010A&ARv..18..127B}
}

@ARTICLE{lee+24a,
       author = {{Lee}, W. and {Pillepich}, A. and {ZuHone}, J. and {Nelson}, D. and {Jee}, M.~J. and {Nagai}, D. and {Finner}, K.},
        title = "{Radio relics in massive galaxy cluster mergers in the TNG-Cluster simulation}",
      journal = {\aap},
         year = 2024,
        month = jun,
       volume = {686},
          eid = {A55},
        pages = {A55},
          doi = {10.1051/0004-6361/202348194},
archivePrefix = {arXiv},
       eprint = {2311.06340},
 primaryClass = {astro-ph.GA},
       adsurl = {https://ui.adsabs.harvard.edu/abs/2024A&A...686A..55L}
}

@ARTICLE{vazza+11,
       author = {{Vazza}, F. and {Dolag}, K. and {Ryu}, D. and {Brunetti}, G. and {Gheller}, C. and {Kang}, H. and {Pfrommer}, C.},
        title = "{A comparison of cosmological codes: properties of thermal gas and shock waves in large-scale structures}",
      journal = {\mnras},
         year = 2011,
        month = dec,
       volume = {418},
       number = {2},
        pages = {960-985},
          doi = {10.1111/j.1365-2966.2011.19546.x},
archivePrefix = {arXiv},
       eprint = {1106.2159},
 primaryClass = {astro-ph.CO},
       adsurl = {https://ui.adsabs.harvard.edu/abs/2011MNRAS.418..960V}
}

@ARTICLE{quilis+98,
       author = {{Quilis}, Vicent and {M}, Jos{\'e} and {Ib{\'a}{\~n}ez}, . and {S{\'a}ez}, Diego},
        title = "{On the Role of Shock Waves in Galaxy Cluster Evolution}",
      journal = {\apj},
         year = 1998,
        month = aug,
       volume = {502},
       number = {2},
        pages = {518-530},
          doi = {10.1086/305932},
archivePrefix = {arXiv},
       eprint = {astro-ph/9803059},
 primaryClass = {astro-ph},
       adsurl = {https://ui.adsabs.harvard.edu/abs/1998ApJ...502..518Q}
}

@ARTICLE{miniati+00,
       author = {{Miniati}, Francesco and {Ryu}, Dongsu and {Kang}, Hyesung and {Jones}, T.~W. and {Cen}, Renyue and {Ostriker}, Jeremiah P.},
        title = "{Properties of Cosmic Shock Waves in Large-Scale Structure Formation}",
      journal = {\apj},
         year = 2000,
        month = oct,
       volume = {542},
       number = {2},
        pages = {608-621},
          doi = {10.1086/317027},
archivePrefix = {arXiv},
       eprint = {astro-ph/0005444},
 primaryClass = {astro-ph},
       adsurl = {https://ui.adsabs.harvard.edu/abs/2000ApJ...542..608M}
}

@ARTICLE{hong+14,
       author = {{Hong}, Sungwook E. and {Ryu}, Dongsu and {Kang}, Hyesung and {Cen}, Renyue},
        title = "{Shock Waves and Cosmic Ray Acceleration in the Outskirts of Galaxy Clusters}",
      journal = {\apj},
         year = 2014,
        month = apr,
       volume = {785},
       number = {2},
          eid = {133},
        pages = {133},
          doi = {10.1088/0004-637X/785/2/133},
archivePrefix = {arXiv},
       eprint = {1403.1420},
 primaryClass = {astro-ph.CO},
       adsurl = {https://ui.adsabs.harvard.edu/abs/2014ApJ...785..133H}
}

@ARTICLE{skillman+08,
       author = {{Skillman}, Samuel W. and {O'Shea}, Brian W. and {Hallman}, Eric J. and {Burns}, Jack O. and {Norman}, Michael L.},
        title = "{Cosmological Shocks in Adaptive Mesh Refinement Simulations and the Acceleration of Cosmic Rays}",
      journal = {\apj},
         year = 2008,
        month = dec,
       volume = {689},
       number = {2},
        pages = {1063-1077},
          doi = {10.1086/592496},
archivePrefix = {arXiv},
       eprint = {0806.1522},
 primaryClass = {astro-ph},
       adsurl = {https://ui.adsabs.harvard.edu/abs/2008ApJ...689.1063S}
}

@ARTICLE{oshea+04,
       author = {{O'Shea}, Brian W. and {Bryan}, Greg and {Bordner}, James and {Norman}, Michael L. and {Abel}, Tom and {Harkness}, Robert and {Kritsuk}, Alexei},
        title = "{Introducing Enzo, an AMR Cosmology Application}",
      journal = {arXiv e-prints},
         year = 2004,
        month = mar,
          eid = {astro-ph/0403044},
        pages = {astro-ph/0403044},
          doi = {10.48550/arXiv.astro-ph/0403044},
archivePrefix = {arXiv},
       eprint = {astro-ph/0403044},
 primaryClass = {astro-ph},
       adsurl = {https://ui.adsabs.harvard.edu/abs/2004astro.ph..3044O}
}

@ARTICLE{bryan+14,
       author = {{Bryan}, Greg L. and {Norman}, Michael L. and {O'Shea}, Brian W. and {Abel}, Tom and {Wise}, John H. and {Turk}, Matthew J. and {Reynolds}, Daniel R. and {Collins}, David C. and {Wang}, Peng and {Skillman}, Samuel W. and {Smith}, Britton and {Harkness}, Robert P. and {Bordner}, James and {Kim}, Ji-hoon and {Kuhlen}, Michael and {Xu}, Hao and {Goldbaum}, Nathan and {Hummels}, Cameron and {Kritsuk}, Alexei G. and {Tasker}, Elizabeth and {Skory}, Stephen and {Simpson}, Christine M. and {Hahn}, Oliver and {Oishi}, Jeffrey S. and {So}, Geoffrey C. and {Zhao}, Fen and {Cen}, Renyue and {Li}, Yuan and {Enzo Collaboration}},
        title = "{ENZO: An Adaptive Mesh Refinement Code for Astrophysics}",
      journal = {\apjs},
         year = 2014,
        month = apr,
       volume = {211},
       number = {2},
          eid = {19},
        pages = {19},
          doi = {10.1088/0067-0049/211/2/19},
archivePrefix = {arXiv},
       eprint = {1307.2265},
 primaryClass = {astro-ph.IM},
       adsurl = {https://ui.adsabs.harvard.edu/abs/2014ApJS..211...19B}
}

@ARTICLE{vazza+09,
       author = {{Vazza}, F. and {Brunetti}, G. and {Gheller}, C.},
        title = "{Shock waves in Eulerian cosmological simulations: main properties and acceleration of cosmic rays}",
      journal = {\mnras},
         year = 2009,
        month = may,
       volume = {395},
       number = {3},
        pages = {1333-1354},
          doi = {10.1111/j.1365-2966.2009.14691.x},
archivePrefix = {arXiv},
       eprint = {0808.0609},
 primaryClass = {astro-ph},
       adsurl = {https://ui.adsabs.harvard.edu/abs/2009MNRAS.395.1333V}
}

@ARTICLE{keshet+03,
       author = {{Keshet}, Uri and {Waxman}, Eli and {Loeb}, Abraham and {Springel}, Volker and {Hernquist}, Lars},
        title = "{Gamma Rays from Intergalactic Shocks}",
      journal = {\apj},
         year = 2003,
        month = mar,
       volume = {585},
       number = {1},
        pages = {128-150},
          doi = {10.1086/345946},
archivePrefix = {arXiv},
       eprint = {astro-ph/0202318},
 primaryClass = {astro-ph},
       adsurl = {https://ui.adsabs.harvard.edu/abs/2003ApJ...585..128K}
}

@ARTICLE{hoeft+08,
       author = {{Hoeft}, M. and {Br{\"u}ggen}, M. and {Yepes}, G. and {Gottl{\"o}ber}, S. and {Schwope}, A.},
        title = "{Diffuse radio emission from clusters in the MareNostrum Universe simulation}",
      journal = {\mnras},
         year = 2008,
        month = dec,
       volume = {391},
       number = {4},
        pages = {1511-1526},
          doi = {10.1111/j.1365-2966.2008.13955.x},
archivePrefix = {arXiv},
       eprint = {0807.1266},
 primaryClass = {astro-ph},
       adsurl = {https://ui.adsabs.harvard.edu/abs/2008MNRAS.391.1511H}
}

@ARTICLE{pfrommer+06,
       author = {{Pfrommer}, C. and {Springel}, V. and {En{\ss}lin}, T.~A. and {Jubelgas}, M.},
        title = "{Detecting shock waves in cosmological smoothed particle hydrodynamics simulations}",
      journal = {\mnras},
         year = 2006,
        month = mar,
       volume = {367},
       number = {1},
        pages = {113-131},
          doi = {10.1111/j.1365-2966.2005.09953.x},
archivePrefix = {arXiv},
       eprint = {astro-ph/0603483},
 primaryClass = {astro-ph},
       adsurl = {https://ui.adsabs.harvard.edu/abs/2006MNRAS.367..113P}
}

@ARTICLE{hu+14,
       author = {{Hu}, Chia-Yu and {Naab}, Thorsten and {Walch}, Stefanie and {Moster}, Benjamin P. and {Oser}, Ludwig},
        title = "{SPHGal: smoothed particle hydrodynamics with improved accuracy for galaxy simulations}",
      journal = {\mnras},
         year = 2014,
        month = sep,
       volume = {443},
       number = {2},
        pages = {1173-1191},
          doi = {10.1093/mnras/stu1187},
archivePrefix = {arXiv},
       eprint = {1402.1788},
 primaryClass = {astro-ph.CO},
       adsurl = {https://ui.adsabs.harvard.edu/abs/2014MNRAS.443.1173H}
}

@ARTICLE{schaal+15,
       author = {{Schaal}, Kevin and {Springel}, Volker},
        title = "{Shock finding on a moving mesh - I. Shock statistics in non-radiative cosmological simulations}",
      journal = {\mnras},
         year = 2015,
        month = feb,
       volume = {446},
       number = {4},
        pages = {3992-4007},
          doi = {10.1093/mnras/stu2386},
archivePrefix = {arXiv},
       eprint = {1407.4117},
 primaryClass = {astro-ph.CO},
       adsurl = {https://ui.adsabs.harvard.edu/abs/2015MNRAS.446.3992S}
}

@ARTICLE{schaal+16,
       author = {{Schaal}, Kevin and {Springel}, Volker and {Pakmor}, R{\"u}diger and {Pfrommer}, Christoph and {Nelson}, Dylan and {Vogelsberger}, Mark and {Genel}, Shy and {Pillepich}, Annalisa and {Sijacki}, Debora and {Hernquist}, Lars},
        title = "{Shock finding on a moving-mesh - II. Hydrodynamic shocks in the Illustris universe}",
      journal = {\mnras},
         year = 2016,
        month = oct,
       volume = {461},
       number = {4},
        pages = {4441-4465},
          doi = {10.1093/mnras/stw1587},
archivePrefix = {arXiv},
       eprint = {1604.07401},
 primaryClass = {astro-ph.CO},
       adsurl = {https://ui.adsabs.harvard.edu/abs/2016MNRAS.461.4441S}
}

@ARTICLE{sod+78,
       author = {{Sod}, G.~A.},
        title = "{Review. A Survey of Several Finite Difference Methods for Systems of Nonlinear Hyperbolic Conservation Laws}",
      journal = {Journal of Computational Physics},
         year = 1978,
        month = apr,
       volume = {27},
       number = {1},
        pages = {1-31},
          doi = {10.1016/0021-9991(78)90023-2},
       adsurl = {https://ui.adsabs.harvard.edu/abs/1978JCoPh..27....1S}
}

@ARTICLE{arth+19,
       author = {{Arth}, Alexander and {Donnert}, Julius and {Steinwandel}, Ulrich and {B{\"o}ss}, Ludwig and {Halbesma}, Timo and {P{\"u}tz}, Martin and {Hubber}, David and {Dolag}, Klaus},
        title = "{WVTICs -- SPH initial conditions for everyone}",
      journal = {arXiv e-prints},
         year = 2019,
        month = jul,
          eid = {arXiv:1907.11250},
        pages = {arXiv:1907.11250},
          doi = {10.48550/arXiv.1907.11250},
archivePrefix = {arXiv},
       eprint = {1907.11250},
 primaryClass = {astro-ph.IM},
       adsurl = {https://ui.adsabs.harvard.edu/abs/2019arXiv190711250A}
}

@article{Gaburov&Nitadori2011,
  title = {Astrophysical {{Weighted Particle Magnetohydrodynamics}}},
  author = {Gaburov, Evghenii and Nitadori, Keigo},
  year = 2011,
  month = jun,
  journal = {Mon. Not. R. Astron. Soc.},
  volume = {414},
  number = {1},
  eprint = {1006.4159},
  pages = {129--154},
  issn = {00358711},
  doi = {10.1111/j.1365-2966.2011.18313.x},
  urldate = {2022-01-26},
  archiveprefix = {arXiv}
}

@article{Hopkins2015,
  title = {{{GIZMO}}: {{A New Class}} of {{Accurate}}, {{Mesh-Free Hydrodynamic Simulation Methods}}},
  shorttitle = {{{GIZMO}}},
  author = {Hopkins, Philip F.},
  year = 2015,
  month = jun,
  journal = {Mon. Not. R. Astron. Soc.},
  volume = {450},
  number = {1},
  eprint = {1409.7395},
  pages = {53--110},
  issn = {0035-8711, 1365-2966},
  doi = {10.1093/mnras/stv195},
  urldate = {2022-01-14},
  archiveprefix = {arXiv}
}

@article{Lanson&Vila2008,
  title = {Renormalized {{Meshfree Schemes II}}: {{Convergence}} for {{Scalar Conservation Laws}}},
  shorttitle = {Renormalized {{Meshfree Schemes II}}},
  author = {Lanson, Nathalie and Vila, Jean-Paul},
  year = 2008,
  month = jan,
  journal = {SIAM J. Numer. Anal.},
  volume = {46},
  number = {4},
  pages = {1935--1964},
  publisher = {{Society for Industrial and Applied Mathematics}},
  issn = {0036-1429},
  doi = {10.1137/S003614290444739X},
  urldate = {2022-08-10}
}

@article{Lanson&Vila2008a,
  title = {Renormalized {{Meshfree Schemes I}}: {{Consistency}}, {{Stability}}, and {{Hybrid Methods}} for {{Conservation Laws}}},
  shorttitle = {Renormalized {{Meshfree Schemes I}}},
  author = {Lanson, Nathalie and Vila, Jean-Paul},
  year = 2008,
  month = jan,
  journal = {SIAM J. Numer. Anal.},
  volume = {46},
  number = {4},
  pages = {1912--1934},
  publisher = {{Society for Industrial and Applied Mathematics}},
  issn = {0036-1429},
  doi = {10.1137/S0036142903427718},
  urldate = {2022-08-10}
}

@article{Monaghan&Lattanzio1985,
  title = {A Refined Particle Method for Astrophysical Problems},
  author = {Monaghan, J. J. and Lattanzio, J. C.},
  year = 1985,
  month = aug,
  journal = {Astron. Astrophys.},
  volume = {149},
  number = {1},
  pages = {135--143},
  issn = {0004-6361},
  urldate = {2022-08-12},
  langid = {english}
}

@article{Bezanson2017,
  title={Julia: A fresh approach to numerical computing},
  author={Bezanson, Jeff and Edelman, Alan and Karpinski, Stefan and Shah, Viral B},
  journal={SIAM review},
  volume={59},
  number={1},
  pages={65--98},
  year={2017},
  publisher={SIAM},
  url={https://doi.org/10.1137/141000671}
}

@ARTICLE{Dolag2025,
       author = {{Dolag}, Klaus and {Remus}, Rhea-Silvia and {Valenzuela}, Lucas M. and {Kimmig}, Lucas C. and {Seidel}, Benjamin and {Fortune}, Silvio and {Stoiber}, Johannes and {Ivleva}, Anna and {Hoffmann}, Tadziu and {Biffi}, Veronica and {Marini}, Ilaria and {Popesso}, Paola and {Vladutescu-Zopp}, Stephan},
        title = "{Encyclopedia Magneticum: Scaling Relations from Cosmic Dawn to Present Day}",
      journal = {arXiv e-prints},
         year = 2025,
        month = apr,
          eid = {arXiv:2504.01061},
        pages = {arXiv:2504.01061},
          doi = {10.48550/arXiv.2504.01061},
archivePrefix = {arXiv},
       eprint = {2504.01061},
 primaryClass = {astro-ph.CO},
       adsurl = {https://ui.adsabs.harvard.edu/abs/2025arXiv250401061D}
}

@software{gadgetio,
       author = {{B{\"o}ss}, Ludwig M. and {Valenzuela}, Lucas M.},
        title = "{LudwigBoess/GadgetIO.jl: v0.8.2}",
         year = 2025,
        month = nov,
          eid = {10.5281/zenodo.17523824},
          doi = {10.5281/zenodo.17523824},
      version = {v0.8.2},
    publisher = {Zenodo},
       adsurl = {https://ui.adsabs.harvard.edu/abs/2025zndo..17523824B}
}

@software{sphtogrid,
  author       = {{B{\"o}ss}, Ludwig M.},
  title        = {LudwigBoess/SPHtoGrid.jl: v0.5.3},
  month        = dec,
  year         = 2025,
  publisher    = {Zenodo},
  version      = {v0.5.3},
  doi          = {10.5281/zenodo.17953700},
  url          = {https://doi.org/10.5281/zenodo.17953700},
}

@ARTICLE{McKee1977,
       author = {{McKee}, C.~F. and {Ostriker}, J.~P.},
        title = "{A theory of the interstellar medium: three components regulated by supernova explosions in an inhomogeneous substrate.}",
      journal = {\apj},
         year = 1977,
        month = nov,
       volume = {218},
        pages = {148-169},
          doi = {10.1086/155667},
       adsurl = {https://ui.adsabs.harvard.edu/abs/1977ApJ...218..148M}
}

@ARTICLE{Joung2006,
       author = {{Joung}, M.~K. Ryan and {Mac Low}, Mordecai-Mark},
        title = "{Turbulent Structure of a Stratified Supernova-driven Interstellar Medium}",
      journal = {\apj},
         year = 2006,
        month = dec,
       volume = {653},
       number = {2},
        pages = {1266-1279},
          doi = {10.1086/508795},
archivePrefix = {arXiv},
       eprint = {astro-ph/0601005},
 primaryClass = {astro-ph},
       adsurl = {https://ui.adsabs.harvard.edu/abs/2006ApJ...653.1266J}
}

@ARTICLE{Li2015,
       author = {{Li}, Miao and {Ostriker}, Jeremiah P. and {Cen}, Renyue and {Bryan}, Greg L. and {Naab}, Thorsten},
        title = "{Supernova Feedback and the Hot Gas Filling Fraction of the Interstellar Medium}",
      journal = {\apj},
         year = 2015,
        month = nov,
       volume = {814},
       number = {1},
          eid = {4},
        pages = {4},
          doi = {10.1088/0004-637X/814/1/4},
archivePrefix = {arXiv},
       eprint = {1506.07180},
 primaryClass = {astro-ph.GA},
       adsurl = {https://ui.adsabs.harvard.edu/abs/2015ApJ...814....4L}
}

@ARTICLE{Walch2015,
       author = {{Walch}, S. and {Girichidis}, P. and {Naab}, T. and {Gatto}, A. and {Glover}, S.~C.~O. and {W{\"u}nsch}, R. and {Klessen}, R.~S. and {Clark}, P.~C. and {Peters}, T. and {Derigs}, D. and {Baczynski}, C.},
        title = "{The SILCC (SImulating the LifeCycle of molecular Clouds) project - I. Chemical evolution of the supernova-driven ISM}",
      journal = {\mnras},
         year = 2015,
        month = nov,
       volume = {454},
       number = {1},
        pages = {238-268},
          doi = {10.1093/mnras/stv1975},
archivePrefix = {arXiv},
       eprint = {1412.2749},
 primaryClass = {astro-ph.GA},
       adsurl = {https://ui.adsabs.harvard.edu/abs/2015MNRAS.454..238W}
}

@ARTICLE{Girichidis2016,
       author = {{Girichidis}, Philipp and {Walch}, Stefanie and {Naab}, Thorsten and {Gatto}, Andrea and {W{\"u}nsch}, Richard and {Glover}, Simon C.~O. and {Klessen}, Ralf S. and {Clark}, Paul C. and {Peters}, Thomas and {Derigs}, Dominik and {Baczynski}, Christian},
        title = "{The SILCC (SImulating the LifeCycle of molecular Clouds) project - II. Dynamical evolution of the supernova-driven ISM and the launching of outflows}",
      journal = {\mnras},
         year = 2016,
        month = mar,
       volume = {456},
       number = {4},
        pages = {3432-3455},
          doi = {10.1093/mnras/stv2742},
archivePrefix = {arXiv},
       eprint = {1508.06646},
 primaryClass = {astro-ph.GA},
       adsurl = {https://ui.adsabs.harvard.edu/abs/2016MNRAS.456.3432G}
}

@ARTICLE{Kim2017,
       author = {{Kim}, Chang-Goo and {Ostriker}, Eve C.},
        title = "{Three-phase Interstellar Medium in Galaxies Resolving Evolution with Star Formation and Supernova Feedback (TIGRESS): Algorithms, Fiducial Model, and Convergence}",
      journal = {\apj},
         year = 2017,
        month = sep,
       volume = {846},
       number = {2},
          eid = {133},
        pages = {133},
          doi = {10.3847/1538-4357/aa8599},
archivePrefix = {arXiv},
       eprint = {1612.03918},
 primaryClass = {astro-ph.GA},
       adsurl = {https://ui.adsabs.harvard.edu/abs/2017ApJ...846..133K}
}

@ARTICLE{MacLow1999,
       author = {{Mac Low}, Mordecai-Mark and {Ferrara}, Andrea},
        title = "{Starburst-driven Mass Loss from Dwarf Galaxies: Efficiency and Metal Ejection}",
      journal = {\apj},
         year = 1999,
        month = mar,
       volume = {513},
       number = {1},
        pages = {142-155},
          doi = {10.1086/306832},
archivePrefix = {arXiv},
       eprint = {astro-ph/9801237},
 primaryClass = {astro-ph},
       adsurl = {https://ui.adsabs.harvard.edu/abs/1999ApJ...513..142M}
}

@ARTICLE{Kim2020,
       author = {{Kim}, Chang-Goo and {Ostriker}, Eve C. and {Somerville}, Rachel S. and {Bryan}, Greg L. and {Fielding}, Drummond B. and {Forbes}, John C. and {Hayward}, Christopher C. and {Hernquist}, Lars and {Pandya}, Viraj},
        title = "{First Results from SMAUG: Characterization of Multiphase Galactic Outflows from a Suite of Local Star-forming Galactic Disk Simulations}",
      journal = {\apj},
         year = 2020,
        month = sep,
       volume = {900},
       number = {1},
          eid = {61},
        pages = {61},
          doi = {10.3847/1538-4357/aba962},
archivePrefix = {arXiv},
       eprint = {2006.16315},
 primaryClass = {astro-ph.GA},
       adsurl = {https://ui.adsabs.harvard.edu/abs/2020ApJ...900...61K}
}

@ARTICLE{Kolborg2022,
       author = {{Kolborg}, Anne Noer and {Martizzi}, Davide and {Ramirez-Ruiz}, Enrico and {Pfister}, Hugo and {Sakari}, Charli and {Wechsler}, Risa H. and {Soares-Furtado}, Melinda},
        title = "{Supernova-driven Turbulent Metal Mixing in High-redshift Galactic Disks: Metallicity Fluctuations in the Interstellar Medium and its Imprints on Metal-poor Stars in the Milky Way}",
      journal = {\apjl},
         year = 2022,
        month = sep,
       volume = {936},
       number = {2},
          eid = {L26},
        pages = {L26},
          doi = {10.3847/2041-8213/ac8c98},
archivePrefix = {arXiv},
       eprint = {2111.02619},
 primaryClass = {astro-ph.GA},
       adsurl = {https://ui.adsabs.harvard.edu/abs/2022ApJ...936L..26K}
}

@ARTICLE{Steinwandel2025,
       author = {{Steinwandel}, Ulrich P. and {Rennehan}, Douglas and {Orr}, Matthew E. and {Fielding}, Drummond B. and {Kim}, Chang-Goo},
        title = "{Pumping Iron: How Turbulent Metal Diffusion Impacts Multiphase Galactic Outflows}",
      journal = {\apj},
         year = 2025,
        month = sep,
       volume = {991},
       number = {1},
          eid = {16},
        pages = {16},
          doi = {10.3847/1538-4357/adf283},
archivePrefix = {arXiv},
       eprint = {2407.14599},
 primaryClass = {astro-ph.GA},
       adsurl = {https://ui.adsabs.harvard.edu/abs/2025ApJ...991...16S}
}

@ARTICLE{Balsara2004,
       author = {{Balsara}, Dinshaw S. and {Kim}, Jongsoo and {Mac Low}, Mordecai-Mark and {Mathews}, Grant J.},
        title = "{Amplification of Interstellar Magnetic Fields by Supernova-driven Turbulence}",
      journal = {\apj},
         year = 2004,
        month = dec,
       volume = {617},
       number = {1},
        pages = {339-349},
          doi = {10.1086/425297},
archivePrefix = {arXiv},
       eprint = {astro-ph/0403660},
 primaryClass = {astro-ph},
       adsurl = {https://ui.adsabs.harvard.edu/abs/2004ApJ...617..339B}
}

@ARTICLE{Gressel2008,
       author = {{Gressel}, O. and {Elstner}, D. and {Ziegler}, U. and {R{\"u}diger}, G.},
        title = "{Direct simulations of a supernova-driven galactic dynamo}",
      journal = {\aap},
         year = 2008,
        month = aug,
       volume = {486},
       number = {3},
        pages = {L35-L38},
          doi = {10.1051/0004-6361:200810195},
archivePrefix = {arXiv},
       eprint = {0805.2616},
 primaryClass = {astro-ph},
       adsurl = {https://ui.adsabs.harvard.edu/abs/2008A&A...486L..35G}
}

@ARTICLE{Rieder2016,
       author = {{Rieder}, Michael and {Teyssier}, Romain},
        title = "{A small-scale dynamo in feedback-dominated galaxies as the origin of cosmic magnetic fields - I. The kinematic phase}",
      journal = {\mnras},
         year = 2016,
        month = apr,
       volume = {457},
       number = {2},
        pages = {1722-1738},
          doi = {10.1093/mnras/stv2985},
archivePrefix = {arXiv},
       eprint = {1506.00849},
 primaryClass = {astro-ph.GA},
       adsurl = {https://ui.adsabs.harvard.edu/abs/2016MNRAS.457.1722R}
}

@ARTICLE{Gent2021,
       author = {{Gent}, Frederick A. and {Mac Low}, Mordecai-Mark and {K{\"a}pyl{\"a}}, Maarit J. and {Singh}, Nishant K.},
        title = "{Small-scale Dynamo in Supernova-driven Interstellar Turbulence}",
      journal = {\apjl},
         year = 2021,
        month = apr,
       volume = {910},
       number = {2},
          eid = {L15},
        pages = {L15},
          doi = {10.3847/2041-8213/abed59},
archivePrefix = {arXiv},
       eprint = {2010.01833},
 primaryClass = {astro-ph.GA},
       adsurl = {https://ui.adsabs.harvard.edu/abs/2021ApJ...910L..15G}
}

@ARTICLE{Brueggen2012,
       author = {{Br{\"u}ggen}, Marcus and {Bykov}, Andrei and {Ryu}, Dongsu and {R{\"o}ttgering}, Huub},
        title = "{Magnetic Fields, Relativistic Particles, and Shock Waves in Cluster Outskirts}",
      journal = {\ssr},
         year = 2012,
        month = may,
       volume = {166},
       number = {1-4},
        pages = {187-213},
          doi = {10.1007/s11214-011-9785-9},
archivePrefix = {arXiv},
       eprint = {1107.5223},
 primaryClass = {astro-ph.HE},
       adsurl = {https://ui.adsabs.harvard.edu/abs/2012SSRv..166..187B}
}

@ARTICLE{Bykov2008,
       author = {{Bykov}, A.~M. and {Dolag}, K. and {Durret}, F.},
        title = "{Cosmological Shock Waves}",
      journal = {\ssr},
         year = 2008,
        month = feb,
       volume = {134},
       number = {1-4},
        pages = {119-140},
          doi = {10.1007/s11214-008-9312-9},
archivePrefix = {arXiv},
       eprint = {0801.0995},
 primaryClass = {astro-ph},
       adsurl = {https://ui.adsabs.harvard.edu/abs/2008SSRv..134..119B}
}

@ARTICLE{Jubelgas2004,
       author = {{Jubelgas}, Martin and {Springel}, Volker and {Dolag}, Klaus},
        title = "{Thermal conduction in cosmological SPH simulations}",
      journal = {\mnras},
         year = 2004,
        month = jun,
       volume = {351},
       number = {2},
        pages = {423-435},
          doi = {10.1111/j.1365-2966.2004.07801.x},
archivePrefix = {arXiv},
       eprint = {astro-ph/0401456},
 primaryClass = {astro-ph},
       adsurl = {https://ui.adsabs.harvard.edu/abs/2004MNRAS.351..423J}
}

@ARTICLE{Arth2014,
       author = {{Arth}, Alexander and {Dolag}, Klaus and {Beck}, Alexander M. and {Petkova}, Margarita and {Lesch}, Harald},
        title = "{Anisotropic thermal conduction in galaxy clusters with MHD in Gadget}",
      journal = {arXiv e-prints},
         year = 2014,
        month = dec,
          eid = {arXiv:1412.6533},
        pages = {arXiv:1412.6533},
          doi = {10.48550/arXiv.1412.6533},
archivePrefix = {arXiv},
       eprint = {1412.6533},
 primaryClass = {astro-ph.CO},
       adsurl = {https://ui.adsabs.harvard.edu/abs/2014arXiv1412.6533A}
}

@ARTICLE{Cullen2010,
       author = {{Cullen}, Lee and {Dehnen}, Walter},
        title = "{Inviscid smoothed particle hydrodynamics}",
      journal = {\mnras},
         year = 2010,
        month = oct,
       volume = {408},
       number = {2},
        pages = {669-683},
          doi = {10.1111/j.1365-2966.2010.17158.x},
archivePrefix = {arXiv},
       eprint = {1006.1524},
 primaryClass = {astro-ph.IM},
       adsurl = {https://ui.adsabs.harvard.edu/abs/2010MNRAS.408..669C}
}

@ARTICLE{Dehnen2012,
       author = {{Dehnen}, Walter and {Aly}, Hossam},
        title = "{Improving convergence in smoothed particle hydrodynamics simulations without pairing instability}",
      journal = {\mnras},
         year = 2012,
        month = sep,
       volume = {425},
       number = {2},
        pages = {1068-1082},
          doi = {10.1111/j.1365-2966.2012.21439.x},
archivePrefix = {arXiv},
       eprint = {1204.2471},
 primaryClass = {astro-ph.IM},
       adsurl = {https://ui.adsabs.harvard.edu/abs/2012MNRAS.425.1068D}
}

@ARTICLE{Springel2010,
       author = {{Springel}, Volker},
        title = "{E pur si muove: Galilean-invariant cosmological hydrodynamical simulations on a moving mesh}",
      journal = {\mnras},
         year = 2010,
        month = jan,
       volume = {401},
       number = {2},
        pages = {791-851},
          doi = {10.1111/j.1365-2966.2009.15715.x},
archivePrefix = {arXiv},
       eprint = {0901.4107},
 primaryClass = {astro-ph.CO},
       adsurl = {https://ui.adsabs.harvard.edu/abs/2010MNRAS.401..791S}
}

@ARTICLE{Teyssier2002,
       author = {{Teyssier}, R.},
        title = "{Cosmological hydrodynamics with adaptive mesh refinement. A new high resolution code called RAMSES}",
      journal = {\aap},
         year = 2002,
        month = apr,
       volume = {385},
        pages = {337-364},
          doi = {10.1051/0004-6361:20011817},
archivePrefix = {arXiv},
       eprint = {astro-ph/0111367},
 primaryClass = {astro-ph},
       adsurl = {https://ui.adsabs.harvard.edu/abs/2002A&A...385..337T}
}

@ARTICLE{Pais2018,
       author = {{Pais}, M. and {Pfrommer}, C. and {Ehlert}, K. and {Pakmor}, R.},
        title = "{The effect of cosmic ray acceleration on supernova blast wave dynamics}",
      journal = {\mnras},
         year = 2018,
        month = aug,
       volume = {478},
       number = {4},
        pages = {5278-5295},
          doi = {10.1093/mnras/sty1410},
archivePrefix = {arXiv},
       eprint = {1805.00128},
 primaryClass = {astro-ph.HE},
       adsurl = {https://ui.adsabs.harvard.edu/abs/2018MNRAS.478.5278P}
}

@ARTICLE{MacLow2004,
       author = {{Mac Low}, Mordecai-Mark and {Klessen}, Ralf S.},
        title = "{Control of star formation by supersonic turbulence}",
      journal = {Reviews of Modern Physics},
         year = 2004,
        month = jan,
       volume = {76},
       number = {1},
        pages = {125-194},
          doi = {10.1103/RevModPhys.76.125},
archivePrefix = {arXiv},
       eprint = {astro-ph/0301093},
 primaryClass = {astro-ph},
       adsurl = {https://ui.adsabs.harvard.edu/abs/2004RvMP...76..125M}
}

@ARTICLE{Elmegreen2004,
       author = {{Elmegreen}, Bruce G. and {Scalo}, John},
        title = "{Interstellar Turbulence I: Observations and Processes}",
      journal = {\araa},
         year = 2004,
        month = sep,
       volume = {42},
       number = {1},
        pages = {211-273},
          doi = {10.1146/annurev.astro.41.011802.094859},
archivePrefix = {arXiv},
       eprint = {astro-ph/0404451},
 primaryClass = {astro-ph},
       adsurl = {https://ui.adsabs.harvard.edu/abs/2004ARA&A..42..211E}
}

@ARTICLE{Springel2005,
       author = {{Springel}, Volker},
        title = "{The cosmological simulation code GADGET-2}",
      journal = {\mnras},
         year = 2005,
        month = dec,
       volume = {364},
       number = {4},
        pages = {1105-1134},
          doi = {10.1111/j.1365-2966.2005.09655.x},
archivePrefix = {arXiv},
       eprint = {astro-ph/0505010},
 primaryClass = {astro-ph},
       adsurl = {https://ui.adsabs.harvard.edu/abs/2005MNRAS.364.1105S}
}

@software{Anthropic_ClaudeOpus55_2026,
  author = {Anthropic},
  title = {Claude Opus 5.5},
  year = {2026},
  url = {https://www.anthropic.com/claude-opus-5-5},
  version = {5.5}
}
\bibliographystyle{aasjournalv7}





\end{document}